# ARTEMIS Observations of Plasma Waves in Laminar and Perturbed Interplanetary Shocks

**L. A. Davis[1]\*, C. A. Cattell[1], L. B. Wilson III[2], Z. A. Cohen[1], A. W. Breneman[1], E. L. M. Hanson[3]**

[1]116 Church Street S.E., School of Physics and Astronomy, University of Minnesota, Minneapolis, Minnesota, 55455, USA

[2]8800 Greenbelt Rd, NASA Goddard Space Flight Center, Greenbelt, Maryland, 20771, USA.

[3]7 Gauss Way, Space Sciences Laboratory, UC Berkeley, Berkeley, California, 94720-7450, USA

\*Corresponding Author: davi2849@umn.edu

**Abstract**

The 'Acceleration, Reconnection, Turbulence and Electrodynamics of the Moon's Interaction with the Sun' (ARTEMIS) mission provides a unique opportunity to study the structure of interplanetary shocks and the associated generation of plasma waves with frequencies between ~50-8000 Hz due to its long duration electric and magnetic field burst waveform captures. We compare wave properties and occurrence rates at 11 quasi-perpendicular interplanetary shocks with burst data within 10 minutes (~3200 proton gyroradii upstream, ~1900 downstream) of the shock ramp. A perturbed shock is defined as possessing a large amplitude whistler precursor in the quasi-static magnetic field with an amplitude greater than ⅓ the difference between the upstream and downstream average magnetic field magnitudes; laminar shocks lack these large precursors and have a smooth, step function-like transition. In addition to wave modes previously observed, including ion acoustic, whistler, and electrostatic solitary waves, waves in the ion acoustic frequency range that show rapid temporal frequency change are common. The ramp region of the two laminar shocks with burst data in the ramp contained a wide range of large amplitude wave modes whereas the one perturbed shock with ramp burst data contained no such waves. Energy dissipation through wave-particle interactions is more prominent in these laminar shocks than the perturbed shock. The wave occurrence rates for laminar shocks are higher in the transition region, especially the ramp, than downstream. Perturbed shocks have approximately 2-3 times the wave occurrence rate downstream than laminar shocks.

**1 Introduction**

Early theoretical work on quasi-perpendicular ($\theta_{Bn} \geq 45°$, where $\theta_{Bn}$ is the angle between the upstream magnetic field and shock normal unit vector) collisionless shocks used the structure of the quasi-static magnetic field to classify the shocks as "laminar", "quasi-laminar", "quasi-turbulent", and "turbulent" based on the upstream average fast mode Mach number $M_f$ and plasma beta $\beta$ [see *Greenstadt*, 1985; *Mellott*, 1985, and references therein]. Until recently [*Wilson III et al.*, 2017], laminar shocks were usually thought to be low Mach number ($M_f \lesssim 3$), low beta ($\beta \lesssim 1$), quasi-perpendicular shocks based on both theory and observation [*Galeev and Karpman*, 1963; *Sagdeev*, 1966; *Tidman and Krall*, 1971; *Biskamp*, 1973; *Greenstadt et al.*, 1975; *Mellott and Greenstadt*, 1984; *Mellott*, 1985; *Farris et al.*, 1993]. These have step function-like ramps, with a sharp magnetic field change between the upstream and downstream regions.

Note that laminar shocks may still exhibit upstream or downstream electromagnetic fluctuations [*Gary and Mellott*, 1985].

Turbulent shocks are expected to occur for high Mach number ($M_f \gtrsim 3$) and/or high beta ($\beta \gtrsim 1$) [*Sagdeev*, 1966; *Kennel and Sagdeev*, 1967a,b; *Coroniti*, 1970b; *Formisano and Hedgecock*, 1973a,b; *Formisano et al.*, 1975; *Wilson III et al.*, 2012]. Turbulent shocks generally possess some feature in the ramp region resulting in a non-laminar structure. Such a non-laminar structure could be generated when a whistler precursor has sufficient amplitude, with respect to the amplitude of the shock ramp, to cause perturbations in the ramp structure [*Wilson III et al.*, 2017].

This division between laminar and turbulent has recently been questioned by *Wilson III et al.* [2017], who looked at the structure of low Mach number ($M_f \leq 3$), low beta ($\beta_{up} \leq 1$), quasi-perpendicular ($\theta_{Bn} \geq 45°$) shocks and found that ~78% of these shocks had clear whistler precursors. When the maximum peak-to-peak amplitudes of these precursors $\delta B_{precursor}$ was compared with the difference between the upstream and downstream average magnetic field magnitudes $\Delta B$, the average (median) value of $\delta B_{precursor}/\Delta B$ was ~2.2 (~1.1). With such large normalized amplitudes, doubts have been raised about the traditional classification of these shocks as "laminar." They also note that whistler precursors may be even more common, but could not be resolved without higher (> 11 samples s$^{-1}$) quasi-static magnetic field sampling rates. Whistler precursors have also been observed to be common occurrence over a range of plasma conditions ($M_f$ between 1.2-2.6, $\beta_{up}$ between 1-16) [*Russell et al.*, 2009; *Ramírez Vélez et al.*, 2012; *Kajdič et al.*, 2012].

Whistler precursors are a manifestation of dispersive radiation [*Tidman and Northrop*, 1968; *Fairfield et al.*, 1974; *Mellott and Greenstadt*, 1984], one mechanism through which collisionless shocks may transform bulk flow kinetic energy into other forms of energy. Several other mechanisms have been proposed, including wave-particle interactions [*Sagdeev, 1966; Coroniti*, 1970a; *Gary*, 1981], particle reflection [*Edmiston and Kennel*, 1984; *Kennel et al.*, 1985; *Kennel*, 1987; *Bale et al.*, 2005; *Su et al.*, 2012], and macroscopic quasi-static field effects [*Scudder et al.*, 1986a,b,c]. At low $M_f$, theory suggests that dispersive radiation and wave-particles interactions dominate [*Kennel et al.*, 1985]. Because this study focuses on low Mach number shocks ($M_f \lesssim 3$), particle reflection and macroscopic field effects are not investigated in detail. Whistler precursors [*Tidman and Northrop*, 1968; *Fairfield et al.*, 1974] are often observed and can dissipate energy through several processes, including generation of higher frequency waves by creating electron temperature anisotropies or current-driven instabilities [*Gary*, 1981; *Hull et al.*, 2012], acceleration of halo electrons and thermal ions [*Wilson III et al.*, 2012; *Chen et al.*, 2018], and deflection and modulation of core particles [*Goncharov et al.*, 2014]. Wave-particle interactions dissipate energy through anomalous resistivity, shorthand for the energy and momentum exchange between the electromagnetic fields and particles [*Sagdeev, 1966; Coroniti*, 1970a; *Gary*, 1981; *Papadopoulos*, 1985; *Breneman et al.*, 2013; *Wilson III et al.*, 2007, 2010, 2012, 2014a,b]. *Goodrich et al.* [2018] showed that ion acoustic waves may be indicators or facilitators of momentum transfer between reflected and incident ion populations, thus linking these waves to the transformation of bulk flow kinetic energy. In addition, work by *Wang et al.* [2020] shows how reflected ions interacting with incident ions generate Debye-scale electrostatic fluctuations, starting as ion acoustic waves that trap ions and decay into electrostatic solitary waves. *Krasnoselskikh et al.* [2013] provides a review of the quasi-perpendicular bow shock and the inferred dissipation mechanisms based on Cluster observations. Some of the types of waves observed near shocks include: magnetosonic whistler mode waves [*Fairfield et al.*, 1974; *Coroniti et al.*, 1982], solitary waves [*Bale et al.*, 1998], ion acoustic waves [*Fredricks et al.*, 1968; *Rodriguez and Gurnett*, 1975; *Gary et al.*, 1975; *Gurnett et al.*, 1979a,b; *Formisano and Torbert*, 1982; *Fuselier and Gurnett*, 1984; *Balikhin et al.*, 2005; *Hull et al.*, 2006], Langmuir waves [*Filbert and Kellogg,* 1979; *Kellogg*, 2003 and references therein; *Pulupa and Bale*, 2008], and electron cyclotron harmonic waves and waves associated with the electron cyclotron drift instability [*Wilson III et al.*, 2010, 2014a,b; *Breneman et al.*, 2013; *Goodrich et al.*, 2018].

Because *Wilson III et al.* [2017] showed that whistler precursors perturbing the ramp region are a common occurrence for low Mach number shocks, it is important to examine whether there is a difference in the transformation of the bulk flow kinetic energy into other forms between laminar and perturbed shocks. Since most methods of energy transformation result in processes involving wave-particle interactions, one approach to this question is analyzing waves observed near shocks. This gives rise to several questions about the possible role of higher frequency waves: Are there differences between laminar and perturbed interplanetary shocks in the wave modes and amplitudes observed near/within the ramp region? Are there differences in the number and/or duration of wave packets observed? To answer these questions, we analyze ARTEMIS long duration electric and magnetic wave burst captures to examine waves at frequencies of ~50-8000 Hz near the ramps of 11 interplanetary (IP) shocks. We briefly review waves in this frequency range.

Whistler mode waves are electromagnetic, right-hand polarized, and occur in the frequency range between the ion cyclotron and electron cyclotron frequencies [e*Beinroth and Neubauer*, 1981; *Lin et al.*, 1998; *Ramírez Vélez et al.*, 2012]. They have been observed both upstream and downstream of IP shocks [*Fairfield*, 1974; *Coroniti et al.*, 1982; *Russell et al.*, 2009; *Ramírez Vélez et al.*, 2012; *Kajdič et al.*, 2012; *Wilson III et al.*, 2013]. In the solar wind, whistlers have been observed in two different frequency bands, a lower and a higher. Lower frequency whistlers, with frequencies from the ion cyclotron frequency up to the lower hybrid frequency, are often whistler precursors, discussed above [*Fairfield et al.*, 1974; *Hoppe et al.*, 1982, 1983; *Russell and Hoppe*, 1983; *Russell et al.*, 2009; *Ramírez Vélez et al.*, 2012; *Kajdič et al.*, 2012]. The higher frequency band of whistlers are most commonly observed near ~0.1-0.3 $f_{ce}$ [*Breneman et al.*, 2010; *Hull et al.*, 2012; *Wilson III et al.*, 2013; *Giagkiozis et al.*, 2018; *Cattell et al.*, 2020], with frequencies up to the electron cyclotron frequency. These higher frequency, narrowband whistlers can have large amplitudes (some > 40 mV m$^{-1}$, in comparison to most whistler amplitudes near ~5 mV m$^{-1}$), and are usually observed at stream interaction regions (~88% of the time) and sometimes at IP shocks (~17%) [*Breneman et al.*, 2010]. The whistler mode waves observed in the electric and magnetic field burst data we use in this study are of the higher frequency band.

Ion acoustic (IA) waves are electrostatic, are linearly or elliptically polarized, have a rest frame frequency up to the ion plasma frequency, $f_{pi}$, (typically observed at ~1-10 kHz in the solar wind), and travel parallel or obliquely to the ambient magnetic field [*Gurnett et al.*, 1979a,b; *Fuselier and Gurnett*, 1984; *Akimoto and Winske*, 1985; *Hess et al.*, 1998; *Balikhin et al.*, 2005]. These waves are thought to be generated by ion-ion or electron-ion drifts [*Gary et al.*, 1975; *Formisano and Torbert*, 1982; *Fuselier and Gurnett,* 1984; *Goodrich et al.*, 2018, 2019]. A number of studies [*Gurnett et al.*, 1979a,b; *Thomsen et al.*, 1985; *Hess et al.*, 1998; *Wilson III et al.*, 2007, 2014a,b; *Chen et al.*, 2018; *Goodrich et al.*, 2018, 2019] have concluded that ion acoustic waves are important in dissipating energy in low Mach number shocks.

Some waves observed in this study displayed characteristics similar to ion acoustic waves. They are electrostatic, have a peak spacecraft-frame frequency between ~1-10 kHz, are linearly or elliptically polarized, and have the peak power primarily parallel to the ambient magnetic field. However, they also show clear frequency change over time. The observed frequency change, which lasts through the ~10-100 ms duration of a given wave packet, is not easily explained by a change due to Doppler shift. The frequency change is not associated with either a rotation or change in magnitude of the quasi-static magnetic field, sampled at 128 samples s$^{-1}$. The change might be due to changes in density; this cannot be ruled out due to the low sampling rate of density measurements. For the purposes of this paper, these waves are referred to as time-dependent frequency electrostatic waves (TFES). Similar wave behavior has also been observed in a STEREO study of IP shocks [*Cohen et al.*, 2020].

Electron cyclotron harmonic waves (ECH) and waves associated with the electron cyclotron drift instability (ECDI) have some characteristics similar to ion acoustic waves [*Matsumoto and Usui*, 1997; *Usui et al.*, 1999; *Wilson III et al.*, 2010; *Breneman et al.* 2013; *Goodrich et al.*, 2018]. The characteristic that distinguishes ECH waves or waves associated with the ECDI from ion acoustic waves is the presence of integer or half-integer harmonics of the electron cyclotron frequency [*Wilson III et al.*, 2010; *Breneman et al.* 2013; *Goodrich et al.*, 2018]. Other features, such as an asymmetric oscillation about a mean or comma-shaped hodograms [*Wilson III et al.*, 2010; *Breneman et al.* 2013; *Goodrich et al.*, 2018], have also been used to distinguish these modes. However in the absence of clear harmonics, as discussed by *Breneman et al.* [2013], it can be difficult to distinguish between these modes. Due to the typically short wavelength ($\lesssim$ 1 electron cyclotron radius) of cyclotron harmonic waves [*Breneman et al.* 2013], higher frequency harmonics are more Doppler shifted than lower frequency harmonics. While sophisticated techniques [*Giagkiozis et al.*, 2011] are available to make this distinction when harmonics are not present, this type of analysis was beyond the scope of this study. Thus any waveform, including ECH waves and waves associated with the ECDI, exhibiting the characteristics of an ion acoustic mode (i.e., linearly or elliptically polarized, $f \sim$ 1-10 kHz in the spacecraft-frame, mostly parallel or oblique to the magnetic field) was identified as an ion acoustic mode. If the characteristics of a TFES mode (i.e., same as ion acoustic, but with clear frequency change over time) were exhibited, the waveform was identified as a TFES mode.

Electrostatic solitary waves are nonlinear bipolar pulses in the electric field which are parallel to the ambient magnetic field and are often associated with electron beams [*Ergun et al.*, 1998; *Cattell et al.*, 2005; *Franz et al.*, 2005]. Solitary waves have been observed at Earth's bow shock [*Bale et al.*, 1998], IP shocks near 1 AU [*Wilson III et al.*, 2007] and near 8.7 AU [*Williams et al.*, 2005], and as a common occurrence in the lunar wake [*Hutchinson and Malaspina*, 2018]. They have also been observed within the magnetosphere, where they potentially act as a mechanism for energy dissipation and particle energization, in the auroral zone [*Ergun et al.*, 1998], magnetotail and plasma sheet boundary layer [*Matsumoto et al.*, 1994; *Andersson et al.*, 2009], reconnection regions [*Cattell et al.*, 2002, 2005], and the radiation belt [*Mozer et al.*, 2013].

In section 2, we describe the instrumentation and methodology for the identification and selection of IP shocks and the wave modes observed. In section 3, we present one example of a laminar shock and one of a perturbed shock. In section 4, we describe the statistics for the 11 shocks and an analysis of the wave modes observed between different shock types. Discussion and conclusions are presented in section 5 and 6, respectively.

## 2 Instrumentation and Methodology

### 2.1 ARTEMIS

The two ARTEMIS spacecraft entered lunar orbit late in 2011. When the spacecraft are not in Earth's magnetosphere or in the lunar wake, they directly measure the solar wind. On board each spacecraft are instruments to measure the magnetic and electric fields, plasma velocity, temperature, and density, and particle distributions [*Angelopoulos*, 2011]. The instruments that measure the fields are the Electric Field Instrument (EFI), the Flux Gate Magnetometer (FGM), and the Search Coil Magnetometer (SCM). The EFI measures the 3D electric field and waves [*Bonnell et al.*, 2008]. The SCM measures the 3D magnetic field fluctuations and waves [*Roux et al.*, 2008]. This study uses the wave burst mode captures from the EFI and SCM, which have a duration on the order of ~10 s [*McFadden et al.*, 2008] and have a sampling frequency of 16384 and 1024 samples s$^{-1}$, respectively [*Bonnell et al.*, 2008; *Roux et al.*, 2008].

The FGM measures the 3D quasi-static magnetic fields [*Auster et al.*, 2008]. This study used FGM data from either the fast survey or particle burst modes, which have a sampling rate of 4 and 128 samples s$^{-1}$, respectively [*Auster et al.*, 2008]. For the events in panel E of Figure 1 and panels B, E, F, G in Figure 2, the particle burst FGM data contained oscillations on the order of a spin period. To remove this artificial effect, a data cleaning algorithm was applied. This algorithm removes periodic noise signals from time-series data for the "target" frequencies: a fundamental frequency set to correspond to the spin period of the spacecraft (~4.3 s) and its harmonics up to the Nyquist frequency. This algorithm is a form of spectral pre-whitening [*MacDonald,* 1989], a class of algorithms commonly used to remove coherent periodic signals from a more continuous incoherent background. The FGM data was first transformed into complex FFT space. A running median of the FFT magnitude was calculated over a window of 400 magnetic field data points (~3.1 s). For any given frequency, a correction factor was derived: the ratio of the FFT magnitude to the median FFT magnitude. A 1 Hz window surrounded and was centered on each target frequency. At each frequency within this window, the corresponding correction factor was modified by a Gaussian and imposed independently upon the real and imaginary FFT components to drive their values toward the local median, filtering the noise peak away while preserving the local trends in phase and amplitude of the FFT. Once the corrected complex FFT was returned to the time domain, spin effects were no longer visible in the FGM data.

The ElectroStatic Analyzer (ESA) measures the electron and ion distributions over the energy range of several eV to 30 keV for electrons and several eV to 25 keV for ions [*McFadden et al.*, 2008]. Three modes are available: full, reduced, and burst packets. These modes are available when the spacecraft is in slow or fast mode, fast mode, or burst mode, respectively. A full packet has a low time resolution (~1 sample min$^{-1}$) but high angular resolution (~2°); a reduced packet has higher time resolution (~15 samples min$^{-1}$) but lower angular resolution (~22°); burst packets have both the higher time resolution and angular resolution [*McFadden et al.*, 2008]. This study used reduced packets or burst packets for the higher time resolution.

ARTEMIS is the only mission primarily in the solar wind with high resolution particle measurements as well as both the quasi-static and 3D wave electric and magnetic field measurements. For example, STEREO lacks a search coil and Wind cannot transmit three components of both field types for the same time interval. ARTEMIS also provides the longest waveform burst captures (~10 s) commonly taken in the solar wind, enabling observations of the evolution and structure of the entire ramp. In comparison, the longest burst captures on STEREO are 2.1 s; for Wind, the burst captures are typically 17 ms, with the longest being 1.13 s. Note that Cluster and MMS have both made burst measurements throughout the ramps of the quasi-perpendicular bow shock [*Balikhin et al.*, 2005; *Krasnoselskikh et al.*, 2013; *Goodrich et al.*, 2018, 2019]. ARTEMIS has a large database (>130) of observed IP shocks. While other missions with similar or higher quality instruments, such as MMS [*Cohen et al.*, 2019; *Hanson et al.*, 2019], have made observations of IP shocks, the number of IP shocks these missions have observed is small in comparison to ARTEMIS. Thus ARTEMIS provides a unique opportunity to study the structure of IP shocks and the associated wave generation and particle energization.

## 2.2 Shock Identification and Parameters

An initial list of IP shock events observed by ARTEMIS between November 2011 and June 2017 was compiled by finding discontinuities in the magnetic field and the ion velocity and density. Only times when ARTEMIS was outside of both the magnetosphere and the lunar wake were considered. If the discontinuity fulfilled the following criteria [*Lumme et al.*, 2017] on the ratio of downstream to upstream magnetic field, ion density and ion temperature, as well as the change in the velocity magnitude, within the uncertainties (relevant uncertainties listed in Table 1),

$$B_{down}/B_{up} \geq 1.2$$
$$N_{down}/N_{up} \geq 1.2$$
$$T_{i,down}/T_{i,up} \geq 1/1.2$$
$$|\Delta V| = |V_{down} - V_{up}| \geq 20\ km\ s^{-1}$$

it was considered a shock and added to the initial event list. A list of IP shocks observed by the ARTEMIS spacecraft between June 2008 and December 2011 (available at *http://themis.ssl.berkeley.edu/data/themis/events/*) was also used.

Only quasi-perpendicular ($\theta_{Bn} \geq 45°$) shocks with wave burst data within 10 minutes, either upstream or downstream, of the ramp were selected for this study. The 10 minute interval was chosen because some whistler precursors (e.g. Panel A, Figure 1) can last this long. Other studies have found that precursors, while typically only extending within ~30,000 km, could extend up to 100,000 km [*Kajdič et al.*, 2012; *Ramírez Vélez et al.*, 2012]. The 10 minute interval corresponds to an average (median) distance of 214,000 (220,000) km and to an average (median) of 3330 (3200) proton gyroradii ($\rho_{gi}$) upstream and 1930 (1160) $\rho_{gi}$ downstream. The specific distances and number of proton gyroradii are different for each event and depend on local plasma parameters. Note that roughly 56% of the burst capture time analyzed in this study occurred within 500 $\rho_{gi}$ of the ramp and 80% occurred within 1500 $\rho_{gi}$. Due to telemetry limitations, of our initial list of >130 shocks, only eleven shocks fulfilled these requirements. For two of the shocks, both spacecraft had burst data available, and the observations from each spacecraft were considered individually, yielding the 13 observations in the event list in Table 1. Of the 13 events, 6 had burst captures overlapping with the transition region, three laminar and three perturbed. Three events, two laminar and one perturbed, had burst captures covering the entire ramp region. Although 13 events does not provide a statistically significant database to reach conclusions for all low Mach number, quasi-perpendicular IP shocks, it is sufficient to fulfill the goal of this study—to explore the differences in wave activity in the range of ~50-8000 Hz between laminar and perturbed shocks.

To minimize the impact secondary ion populations, such as gyrating and field-aligned ion beams reflected by the shock, have on the ion moment calculations, the core of the solar wind was isolated by utilizing a velocity mask when these secondary ion populations were present, most often in the upstream region [*Wilson III et al.*, 2014a, appendix C, and references therein]. From the solar wind core distribution, the ion velocity, density, and temperature moments were recalculated. This recalculation was made with the same methodology as described by *Wilson III et al.* [2014a]. Such a recalculation was needed because the ESA instrument was originally designed for measuring the hot and slow electron and ion distributions within the magnetosphere, not the fast and cold distributions in the solar wind. *Artemyev et al.* [2018] discusses this in more detail and also provides a second method of recalculating the moments. This second method provides corrected ARTEMIS particle measurements to the OMNI database. The results of both methods are both consistent with Wind and ACE observations The *Wilson III et al.* [2014a] methodology was used over the OMNI data because it provides a higher time resolution (~4 s) than the OMNI data is (~96 s). These recalculated moments, as well as the background magnetic field and electron temperature, were used in the Rankine-Hugoniot (RH) equations to estimate the shock normal and other shock parameters [*Koval and Szabo*, 2008], similar to other studies [*Wilson III et al.*, 2014a,b, 2016; *Kanekal et al.*, 2016]. If there are large amplitude fluctuations near the ramp region, the fluctuations could locally perturb the shock normal and modify the RH equations by affecting the incident flow prior to crossing the shock [*Scholer and Belcher*, 1971]. To avoid this issue, intervals, typically 2-3 minutes, for the upstream and downstream inputs to the RH equations were taken from quiet, undisturbed regions ~1-10 minutes from the shock ramp. The ion velocity, density, and temperature were taken from the corrected ion moments which had a resolution of ~15 samples min$^{-1}$. Statistical uncertainties in the results, which can be significant, are listed in Table 1. Note that this analysis was performed for several different time intervals. The results listed in Table 1 are those which minimized the statistical uncertainty in the output parameters while still producing physically meaningful results (e.g., a forward shock normal directed primarily along the GSE -X-axis).

For four events, analysis found no stable shock solution with $M_f > 1.0$ for any of the standard methods for identifying shocks [*Abraham-Shrauner and Yun*, 1976]. Two of these events, and one satisfying $M_f > 1.0$, did not pass the criteria that $|\Delta V| > 20$ km s$^{-1}$ within the uncertainties after the ion moment corrections were made. This is likely due to inaccuracies because the ARTEMIS detectors were not optimized for solar wind measurements. For these cases, Wind and ACE observations of these shocks (located at L1, whereas ARTEMIS is at 1 AU) were utilized to determine shock parameters and all events passed the above shock criteria. For the five events where there was no stable solution of the RH equations with $M_f > 1.0$ and $|\Delta V| > 20$ km s$^{-1}$ within the uncertainties using the ARTEMIS data, the parameters calculated from Wind were used and are listed in Table 1. While large spacecraft separation perpendicular to the flow of the solar wind has been shown to be associated with sometimes large angular deviations of the calculated shock normal [*Szabo*, 2005; *Koval and Szabo*, 2010], it is not expected that the global structure of shocks evolve dramatically when the spacecraft separation is primarily parallel to the flow of the solar wind, as is the case for Wind and ARTEMIS, especially over a scale of only ~0.01 AU. Thus, we have used Wind-based shock parameters as an estimate. A number of other studies have used Wind similarly and found similar shock parameters between Wind and ARTEMIS [*Möstl et al.*, 2012; *Kanekal et al.*, 2016; *Oliveira and Samsonov*, 2018; *Pope et al.*, 2019]. We do note that shock features, such as ripples along the shock surface [*Terasawa et al.*, 2005; *Neugebauer and Giacalone*, 2005], may evolve from L1 to 1 AU, thus the Wind-based parameters provide only an estimate to the local shock parameters at the time of the ARTEMIS observations.

Table 1 gives the date, satellite (using the THEMIS labels), shock classification, the ratios of the average downstream (subscript d) to upstream (subscript u) magnetic field, ion density, ion and electron temperature, the upstream plasma beta, difference between upstream and downstream ion velocities, fast mode Mach number, $\theta_{Bn}$, and the ratio of the fast mode Mach number to the first critical Mach number [*Edmiston and Kennel*, 1984] for each event. The top (bottom) section of the table lists the shock parameters calculated from ARTEMIS (Wind) observations. Events with burst data overlapping with the transition region are denoted by an asterisk (*); double asterisks (**) denote a burst capture overlapped with the ramp region, and by extension, the transition region as well; supercritical shocks are denoted by a dagger (†). A more complete list of both ARTEMIS and Wind calculated parameters, including all uncertainties, for these events is available in a public repository (*https://doi.org/10.5281/zenodo.4008630*).

ARTEMIS-Based Shock Parameters

| Date (Probe) | Shock Type | $\frac{B_d}{B_u}$ | $\frac{N_d}{N_u}$ | $\frac{T_{i,d}}{T_{i,u}}$ | $\frac{T_{e,d}}{T_{e,u}}$ | $\beta_u$ | $|\Delta V|$ (km s$^{-1}$) | $M_f$ | $\theta_{Bn}$ | $\frac{M_f}{M_{cr}}$ |
|---|---|---|---|---|---|---|---|---|---|---|
| 2009-09-03 (B) | P | 1.58±0.12 | 2.98±0.23 | 6.90±8.67 | 1.05±0.05 | 1.08±0.30 | 19±6 | 1.2±0.3 | 73°±6° | 0.7±0.2 |
| 2013-07-09 (C) | P | 1.70±0.05 | 1.36±0.17 | 3.70±0.40 | 1.25±0.04 | 0.25±0.03 | 32±13 | 1.7±0.3 | 81°±5° | 0.7±0.1 |
| 2015-06-21**† (B) | L | 2.67±0.49 | 4.63±0.17 | 2.16±0.11 | 1.51±0.05 | 6.82±0.55 | 59±3 | 1.9±0.3 | 86°±6° | 1.8±0.3 |
| 2015-06-24† (C) | P | 2.20±0.27 | 1.94±0.31 | 5.52±0.64 | 1.53±0.08 | 0.34±0.09 | 155±6 | 3.2±1.9 | 56°±7° | 1.4±0.9 |
| 2016-04-14**† (C) | P | 1.72±0.11 | 1.73±0.10 | 5.15±0.41 | 1.03±0.03 | 1.07±0.08 | 11±12 | 2.2±0.7 | 71°±6° | 1.3±0.4 |
| 2017-02-24** (B) | L | 1.81±0.15 | 2.39±0.26 | 3.74±0.84 | 1.47±0.20 | 0.39±0.07 | 31±21 | 1.5±0.5 | 76°±12° | 0.7±0.2 |
| 2017-02-24* (C) | L | 1.96±0.18 | 2.12±0.21 | 4.66±1.87 | 1.34±0.09 | 0.46±0.05 | 41±22 | 1.6±0.4 | 72°±8° | 0.8±0.2 |
| 2017-05-20 (C) | P | 2.24±0.36 | 3.81±0.52 | 6.48±1.00 | 1.41±0.14 | 0.35±0.13 | 90±20 | 1.4±0.3 | 52°±17° | 0.6±0.1 |
| Wind-Based Shock Parameters | | | | | | | | | | |
| 2012-02-20 (C) | L | 1.97±0.17 | 2.57±0.23 | 1.80±0.29 | 1.22±0.05 | 0.55±0.06 | 66±14 | 1.9±0.6 | 74°±23° | 0.9±0.3 |
| 2012-12-16 (C) | L | 1.42±0.07 | 1.41±0.09 | 1.55±0.22 | 0.99±0.05 | 0.34±0.02 | 26±7 | 1.2±0.6 | 72°±5° | 0.5±0.3 |
| 2013-05-18 (C) | L | 1.70±0.04 | 1.77±0.11 | 1.58±0.30 | 1.22±0.06 | 0.19±0.01 | 60±5 | 1.5±0.2 | 76°±6° | 0.6±0.1 |
| 2015-06-27* (C) | P | 1.80±0.12 | 3.03±0.72 | 3.09±1.40 | 1.50±0.10 | 0.07±0.02 | 43±17 | 1.2±0.4 | 45°±1° | 0.5±0.2 |
| 2017-05-20*† (B) | P | 2.27±0.49 | 2.57±0.33 | 2.18±0.60 | 1.53±0.07 | 0.37±0.13 | 113±21 | 2.1±0.7 | 74°±12° | 1.0±0.3 |

Table 1: Shock parameters for the 13 events in this study. The top and bottom sections of the table list the shock parameters calculated from ARTEMIS and Wind observations, respectively. Dates with asterisks (*) denote a burst capture overlapped with the transition region; double asterisks (**) denote a burst capture overlapped with the ramp region; daggers (†) denote supercritical shocks. Subscript u (d) represents the average upstream (downstream) value.

### 2.3 Shock Classification

The classification of each shock as laminar or perturbed utilized a definition similar to that of *Wilson III et al.* [2017], where a normalized amplitude between the whistler precursor and the magnetic field jump greater than 10% was considered non-laminar. This definition has been modified to be stricter (10% was changed to 33%),

$$\frac{\delta B_{pre}}{\Delta B} \gtrsim \frac{1}{3}, \Delta B = (B_{up} - B_{down}),$$

where $B_{up}$ and $B_{down}$ are the average upstream and downstream magnetic field magnitudes, respectively, and $\delta B_{pre}$ is the maximum peak-to-peak amplitude of the observed whistler precursor. We refer to an event as perturbed if it had a precursor satisfying this condition. Of the 13 events in this study, 6 were laminar and 7 were perturbed.

The ramp region, the region of the largest change in the magnetic field, for each event was found through visual inspection. Any potential foot, overshoot, or precursor was excluded from the ramp region. The ramp regions found in this study resemble the ramp regions found in other studies [*Hobara et al.*, 2010; *Mazelle et al.*, 2010]. The transition region is the region from the undisturbed upstream to the nominal downstream [*Schwartz and Burgess*, 1991; *Wilson III*, 2016], including any whistler precursor, foot, or overshoot. Note that the ramp region is contained within the transition region.

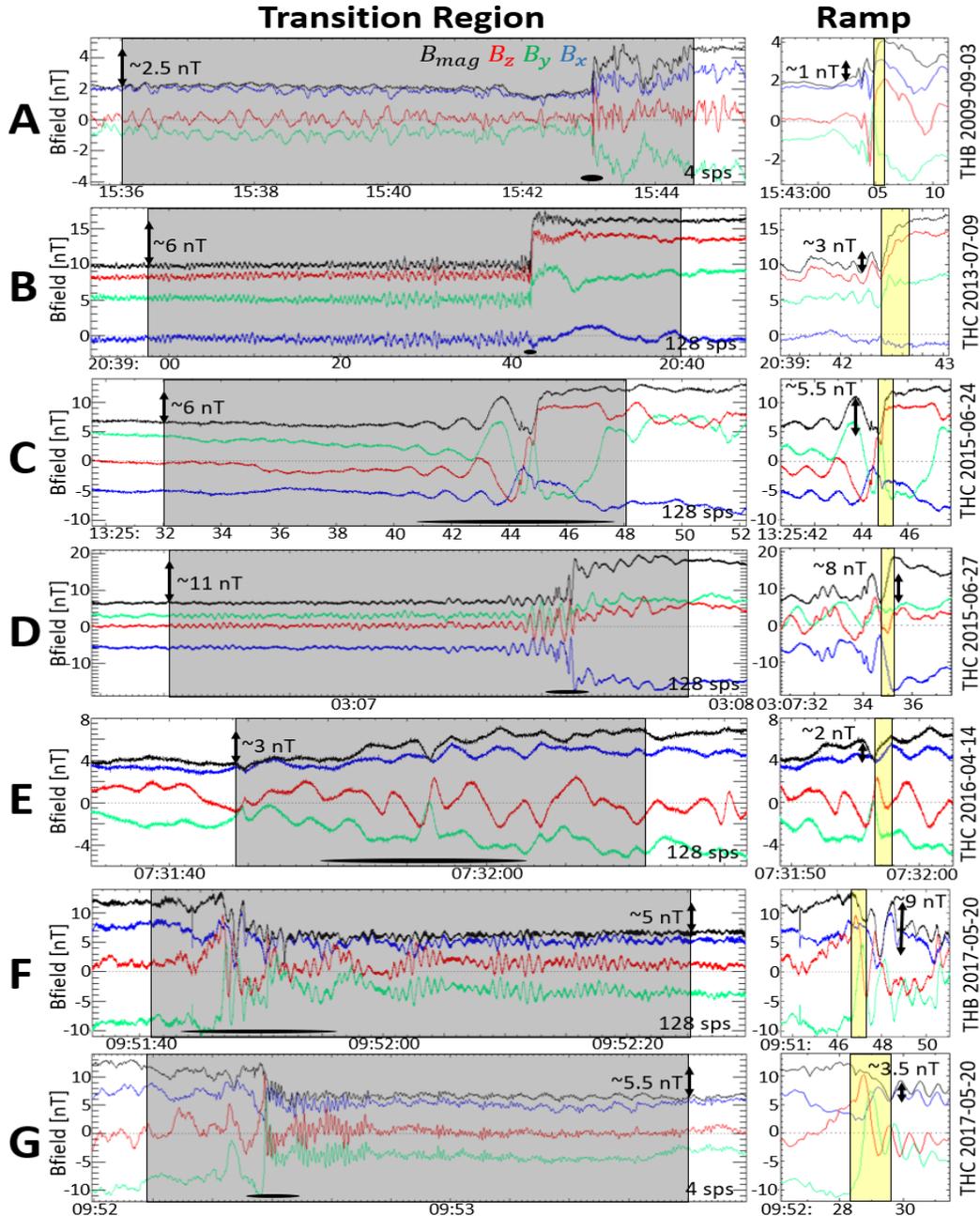

Figure 1: (left) The quasi-static magnetic field magnitude and GSE components near the transition region (shaded in grey) for the 7 perturbed events. The magnitude of the jump from the upstream to the downstream field is labeled in black. The number of samples per second is listed in the lower right. The black oval near the time-axis denotes the zoom-in region to the right. (right) Zoom-in of the magnetic field near the ramp (shaded in yellow). Maximum amplitude of the whistler precursor is labeled in black.

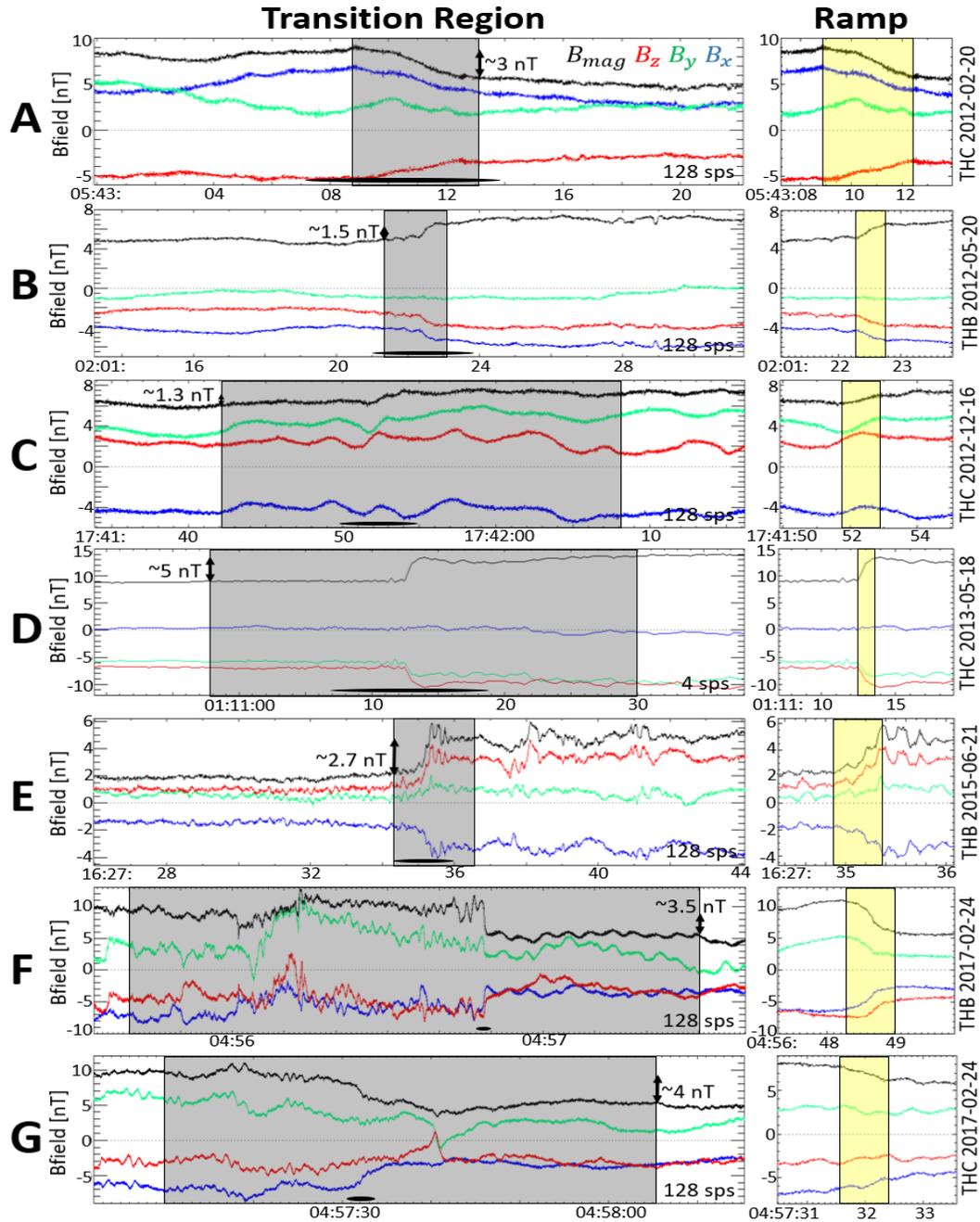

Figure 2: (left) The quasi-static magnetic field magnitude and GSE components near the transition region (shaded in grey) for the 6 laminar events. The magnitude of the jump from the upstream to the downstream field is labeled in black. The number of samples per second is listed in the lower right. The black oval near the time-axis denotes the zoom-in region to the right. (right) Zoom-in of the magnetic field near the ramp (shaded in yellow).

Figure 1 shows the magnetic field magnitude and GSE components for each perturbed event; Figure 2 shows the same quantities for each laminar event. The transition region is shaded grey in the left column; the ramp region is shaded in yellow in the right column. The difference between the upstream and downstream average magnetic field magnitude is shown for each event. Figure 1 also shows the maximum peak-to-peak amplitude of the whistler precursor for each event. The sampling rate of the magnetic field is listed in samples $s^{-1}$ (sps); note that events in panels A and G of Figure 1 and panel D of Figure 2 had a sampling rate of 4 samples $s^{-1}$. Panel D of Figure 2 shows several under resolved fluctuations near the start of the ramp region, but, due to the low sample rate, it is not possible to determine the amplitude of this fluctuation or if it is a whistler precursor. For this reason, the event was classified as laminar.

For low Mach number ($M_f \lesssim 3$), low beta ($\beta_{up} \lesssim 1$) shocks, *Ramírez Vélez et al.* [2012] and *Wilson III et al.* [2017] found no relation between whistler precursors, a feature associated with perturbed shocks, and $\theta_{Bn}$, $M_f$, or any other shock parameter. Whistler precursors are observed at both sub- and super-critical shocks within these range of parameters [*Ramírez Vélez et al.,* 2012; *Wilson III et al.,* 2012]. For these reasons, we have focused this comparative study of plasma waves on laminar and perturbed shocks instead of $\theta_{Bn}$ and $M_f$. All events had $M_f \lesssim 3$; the average $M_f$ (average of ~1.62 for laminar, ~1.86 for perturbed) differed by ~0.24 between the shock types. The average $\theta_{Bn}$ (~76° for laminar; ~64° for perturbed) differed by ~12° between the shock types. All six laminar events had a $\theta_{Bn}$ range from 72-86°; three of seven perturbed events had $\theta_{Bn} < 70°$. Potential bias in our data set based on $M_f$ or $\theta_{Bn}$ is explored in Section 4. The average downstream to upstream electron temperatures did not vary significantly between shock types (average of 1.29 for laminar, 1.33 for perturbed). However, the average downstream to upstream ion temperature ratio for perturbed shocks was roughly twice that for laminar shocks (average of 2.58 for laminar, 4.72 for perturbed).

Properties of the whistler precursors are given in Table 2. Note that the amplitude ratio uses maximum peak-to-peak amplitude of the precursor. The listed frequency is the frequency at peak power found by using Morlet wavelets transforms [*Morlet et al.,* 1982; *Morlet,* 1982]. The range of amplitude ratios (0.4-1.8) and frequencies (0.04-1.55 Hz) is consistent with *Wilson III et al.* [2017]. Our range of frequencies are also consistent with the findings of *Ramírez Vélez et al.* [2012]. By definition, every perturbed event in this study had an associated whistler precursor.

| Date | Probe | $f$ (Hz) | $\frac{\delta B_{pre}}{\Delta B}$ |
|------|-------|----------|------------------------------------|
| 2009-09-03 | B | 0.05 | 0.4 |
| 2013-07-09 | C | 1.25 | 0.5 |
| 2015-06-24 | C | 0.75 | 0.9 |
| 2015-06-27 | C | 0.75 | 0.7 |
| 2016-04-14 | C | 0.25 | 0.7 |
| 2017-05-20 | B | 1.55 | 1.8 |
| 2017-05-20 | C | 1.50 | 0.6 |

Table 2: List of events, all classified as perturbed, with whistler precursors. The frequency $f$ and amplitude with respect to the difference between the upstream and downstream average magnetic field magnitude $\frac{\delta B_{pre}}{\Delta B}$ of the whistler precursor are listed.

## 2.4 Wave Mode Identification

We use the term *wave packet* to refer to a subinterval of a wave burst capture which contains an identifiable wave mode of sufficient power above the background. Our algorithm examined only the electric field power spectrum to find subintervals within each burst capture with power above $10^{-3}$ mV$^2$ m$^{-2}$ Hz$^{-1}$, which were then marked as potential wave packets. Note that due to the power threshold, wave packets with low ($\lesssim 2$ mV m$^{-1}$) electric fields were not included in this study. Although the magnetic field is used in the wave mode identification, it was not used in the initial wave packet identification because of the lower sampling rate. A more detailed description of this algorithm and an example of wave packet identification is in Appendix A. This algorithm can identify multiple packets at different frequencies within the same time interval. Amplitudes stated in this study $|A|$ were calculated using all three field components, $|A| = \sqrt{A_x^2 + A_y^2 + A_z^2}$ where $A_i$ is the $i^{\text{th}}$ component of the $A$ field measured from peak-to-peak. Wave amplitudes observed in this study are large compared to both the motional electric field (~few mV m$^{-1}$) and amplitudes inferred from other studies using spectral data [*Gurnett et al.*, 1979a,b; *Lin* *et al.*, 1998]. All data were rotated into magnetic-field aligned coordinates (FAC), where the *z*-axis was parallel to the ambient magnetic field, and the x- and y-axes were perpendicular to both the ambient magnetic field and to each other.

To identify wave modes, characteristics such as peak frequency, waveform structure, and polarization were used. Figure 3 shows examples of the wave modes observed in this study: whistler, ion acoustic, TFES, and solitary waves. Panels A, B, and C in Figure 3 show the same time interval observed by THC on 2013-07-09. Figure 3a illustrates the process for identifying wave modes from wave packets. These panels show the electric and magnetic field data containing two wave packets identified by the algorithm, each highlighted in blue, and also the surrounding fields. Initial inspection would suggest there are three waves: two with peak frequency ~1000 Hz corresponding to each wave packet and one, at lower amplitudes, with peak frequency ~250 Hz observed in association with both packets and the adjacent intervals. To study these two different wave modes, a bandpass filter was applied. The lower bandpass (3b), from 100 to 400 Hz, showed activity in both the electric and magnetic field. The magnetic field hodograms showed right-handed, nearly circular polarization. The electric field hodograms did not show clear polarization, likely because of contamination by the second, higher frequency wave not removed by the bandpass filter. The peak frequency of ~0.38 $f_{ce}$ and the right-handed magnetic field polarization—identified this wave as a whistler mode wave. The higher bandpass (3c), from 400 to 8000 Hz, coincides roughly with the ion acoustic frequency range of ~1-10 kHz. Hodograms of the electric field showed nearly linear polarization. The highest amplitude component of the electric field was parallel to the magnetic field. The power spectrum of the parallel component of the electric field showed no change in frequency for either wave packet. The peak frequency of ~1 kHz and linear polarization identified these waves as ion acoustic waves. These distinct ion acoustic waves were identified as two waves rather than one because the power dropped below our power threshold for ≳ 50% of either adjacent wave packet duration. Although often each wave packet contained only a single wave mode, sometimes, as in this example, two different wave modes were identified in the same time interval.

For the interval shown in Figure 3d, observed by THC on 2015-06-24, the power spectrum showed the peak frequency was within the ~1-10 kHz range characteristic of ion acoustic waves, but was also increasing with time over an interval of ~0.04 s, so the wave was identified as a TFES wave mode. On 2015-06-27, THC observed short pulses in the electric field. These pulses, with the largest amplitude component parallel to the magnetic field, shown in Figure 3e, were identified as solitary waves. Also shown in the power spectrum is an ion acoustic mode wave, identified from the peak frequency of ~4500 Hz and polarization (not shown). Note that even with the EFI sampling rate of 16384 samples $s^{-1}$, electrostatic solitary waves were often under resolved, so amplitudes listed for this wave mode may be underestimates.

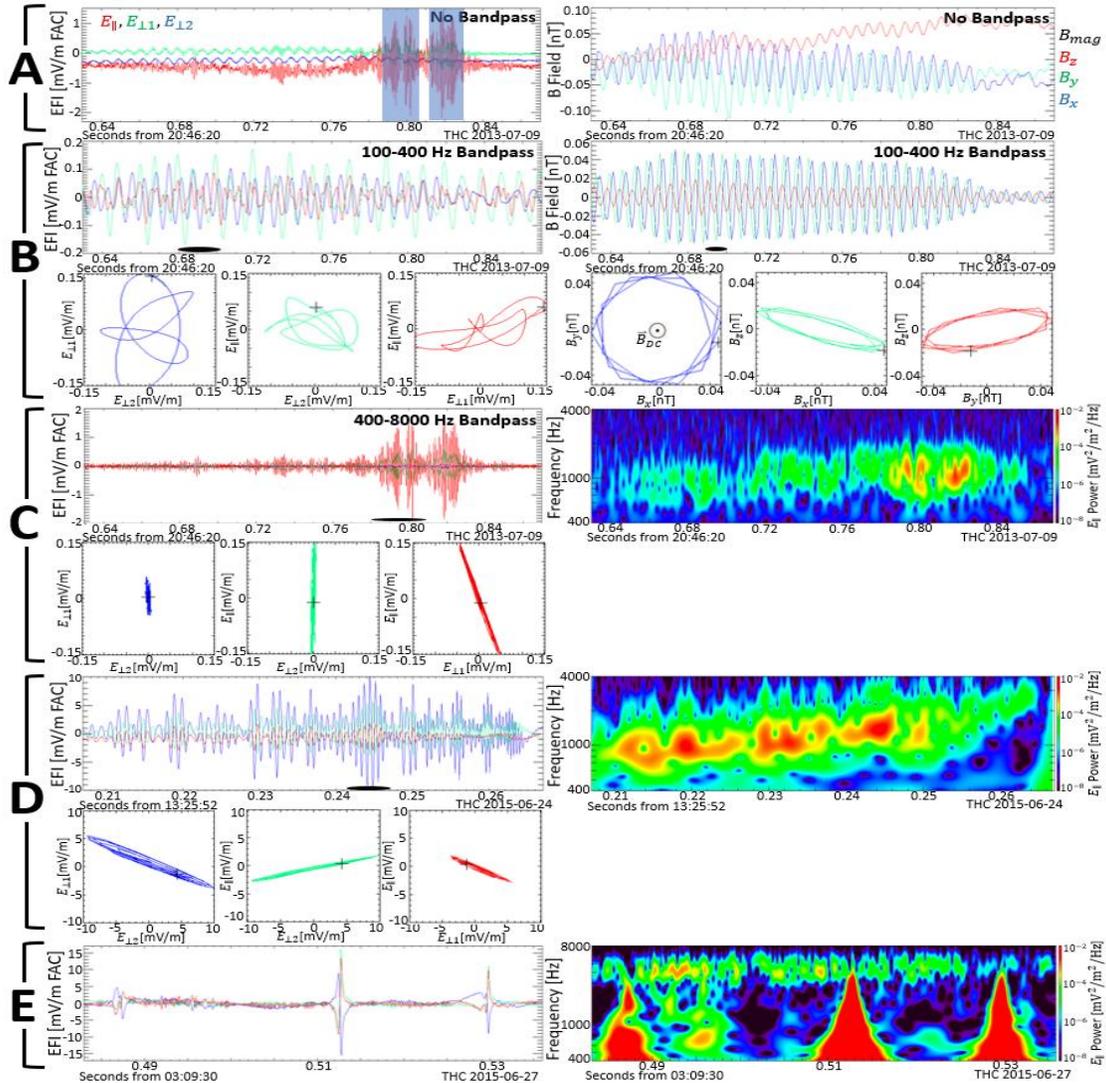

Figure 3: Examples of observed wave modes. (a) Electric (left) and magnetic (right) field showing two wave packets, highlighted in blue. A whistler wave and two ion acoustic waves are observed. (b) An example whistler wave. The electric (left) and magnetic (right) field from A with a bandpass from 100 to 400 Hz applied are shown. (c) Example ion acoustic waves. The electric field (left) from A with a bandpass from 400 to 8000 Hz applied and its power spectrum (right) are shown. (d) An example TFES wave. The electric field (left) and its power spectrum (right) are shown. (e) Example solitary waves. Each pulse in the electric field (left) and power spectrum (right) corresponds to a solitary wave. Hodograms for time intervals with the black bars in the electric and/or magnetic field in B, C, and D are shown beneath their respective plot. The starting point is denoted with a plus.

## 3 Observations

Two illustrative events with burst captures in the transition region are discussed in detail. The 2015-06-21 event was a laminar shock with burst data covering the entire transition region, including the ramp, as well as ~20 s of burst data in the downstream region within 30 seconds (15 $\rho_{gi}$) of the shock ramp. The 2017-05-20 event observed by THB was a perturbed shock with ~10 s of burst data within the transition region, beginning within 1 second (<1 $\rho_{gi}$) from the ramp region. Four other events had burst data within the transition region, for a total of three for each shock classification. For brevity, only two of these six events are discussed, one for each shock type.

### 3.1 Laminar Shock

An overview of the 2015-06-21 shock observed by THB is shown in Figure 4. The ramp occurred at roughly 16:27:35, as indicated by the jump in the magnetic field (4d) and the ion density, velocity, and temperature (4c), as well as electron energy flux enhancements, particularly at lower (~10-200 eV) energies. The electron pitch angle distributions between ~100 eV to 1 keV (4b.1-2) showed a large enhancement for roughly 20 s (~12 $\rho_{gi}$) after the shock across all pitch angles, with the largest increase, antiparallel to the magnetic field, lasting only ~5 s (~3 $\rho_{gi}$). Electrons with these energies were also observed upstream traveling parallel to the magnetic field away from the sun, consistent with strahl. These features were absent in lower energy distributions (4b.3). The parallel component of the electric field (4e) shows three $\gtrsim$ 150 mV m$^{-1}$ peaks in the first burst capture. Only the first peak was associated with a wave mode. The other two peaks were deemed artificial after examination of voltage probe signals.

Figure 5 shows the background magnetic field near the ramp (5a). One burst capture (5b,d) covered the entire ramp, shaded in yellow, and transition region, shaded in grey, and a second burst was taken 30 seconds (15 $\rho_{gi}$) downstream (4e). Because the wave burst magnetic field data started ~1 second after the electric field burst and lacked any significant power (>10$^{-5}$ nT$^2$ Hz$^{-1}$) in the magnetic field burst frequency range of ~30-512 Hz, the particle burst (Nyquist frequency 256 Hz) magnetic field data (5b) and its total power spectrum (5c) are shown instead. The 3D electric burst data (5d) and its total power spectrum (5e) contained more and larger amplitude wave packets than any other burst captures in this study. The electron cyclotron frequency and ion plasma frequency are overplotted on the magnetic and electric power spectra, respectively.

Figure 6 shows the electric field burst waveform (6a) and the associated power spectrum (6b), wave angle with respect to the local, 3-second-averaged magnetic field direction [*Means et al.*, 1972] (6c), and ellipticity [*Samson and Olson*, 1980] (6d) within the ramp region. The shaded regions correspond to different wave packets observed in the ramp. All waves were either linearly or elliptically polarized and had a high degree of polarization. Most waves observed in the ramp had large amplitudes (>20 mV m$^{-1}$) in comparison to the majority of wave packets in this study (see section 4). Region 1 contained an ion acoustic wave, with a peak amplitude of ~169 mV m$^{-1}$ and broadband power from ~50-3000 Hz. The wave angle for this wave packet was the most parallel seen in the ramp, roughly 35° with respect to the local magnetic field. Region 2 contained an ion acoustic wave with an amplitude of ~119 mV m$^{-1}$ and broadband power from ~2500-6500 Hz. Region 3 contained a TFES wave, decreasing in frequency from 5000 Hz to 1000 Hz in ~0.1 seconds, with an amplitude of ~28 mV m$^{-1}$. In region 4, an ion acoustic wave was observed with an amplitude of ~36 mV m$^{-1}$ and peak frequency of ~2000 Hz. Region 5, a zoom-in of which is shown at the bottom of Figure 6, contained a solitary wave with an amplitude ~18 mV m$^{-1}$. Several low amplitude solitary waves occurred in addition to the larger amplitude solitary wave (small black arrows), but because these waves were below our power threshold, they were not included in the statistics. This variability of wave modes and amplitudes observed within the ramp are consistent with the findings of other studies [*Wilson III et al.*, 2007, 2014b; *Breneman et al.*, 2013; *Goodrich et al.*, 2018; *Cohen et al.*, 2020].

These wave modes were also observed outside the ramp region. All waves with amplitudes >20 mV m$^{-1}$ occurred within ~5 seconds (~3 $\rho_{gi}$) of the shock ramp. This proximity to the ramp suggests that the free energy source of these waves is located within or in close proximity to the shock ramp region, consistent with the findings of *Wilson III et al.* [2007]. Two of the waveforms within ~5 seconds (~3 $\rho_{gi}$) of the ramp, shown in Figure 7, are TFES waves and highlight the distinction between this wave mode and ion acoustic waves: the characteristic frequency decreasing (7a) or increasing (7b) with time. The shaded time intervals correspond to the waveforms in the right column of Figure 7. The wave period in T1 is less than that in T2, and the wave period in T3 is greater than that in T4. If the change in frequency observed for the wave in A

(B) was due to Doppler shift, assuming the wave vector was constant because no change in the direction or magnitude of the quasi-static magnetic field was observed, the velocity of the solar wind would need to have changed by a factor >2 in ~0.1 (~0.2) seconds, which was not observed. We examined whether there were harmonics of the electron cyclotron frequency to test for the ECDI. Although the waves had multiple peaks, the peaks were not separated by either integer or half integer multiples of the electron cyclotron frequency, and did not provide evidence for ECH waves or ECDI driven waves.

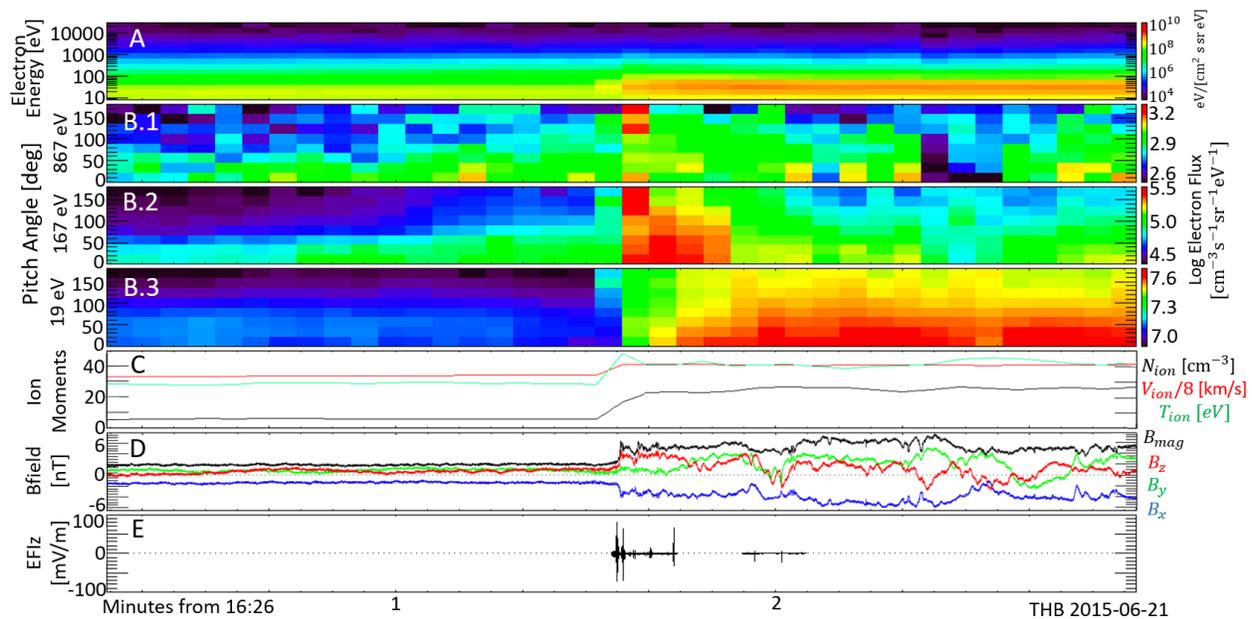

Figure 4: An overview of the 2015-06-21 shock event as observed by THB. (a) Electron energy flux from ~10 eV to 30 keV from ESA. (b.1-3) ESA electron pitch angle distributions for 19 eV, 167 eV, and 867 eV. Color bar is in units of log electron flux. (c) The ion density (cm⁻³, black), velocity (km s⁻¹, red, divided by 8), and temperature (eV, green). (d) Magnetic field in GSE coordinates along with the magnitude (black). (e) Parallel component of the electric field wave burst captures.

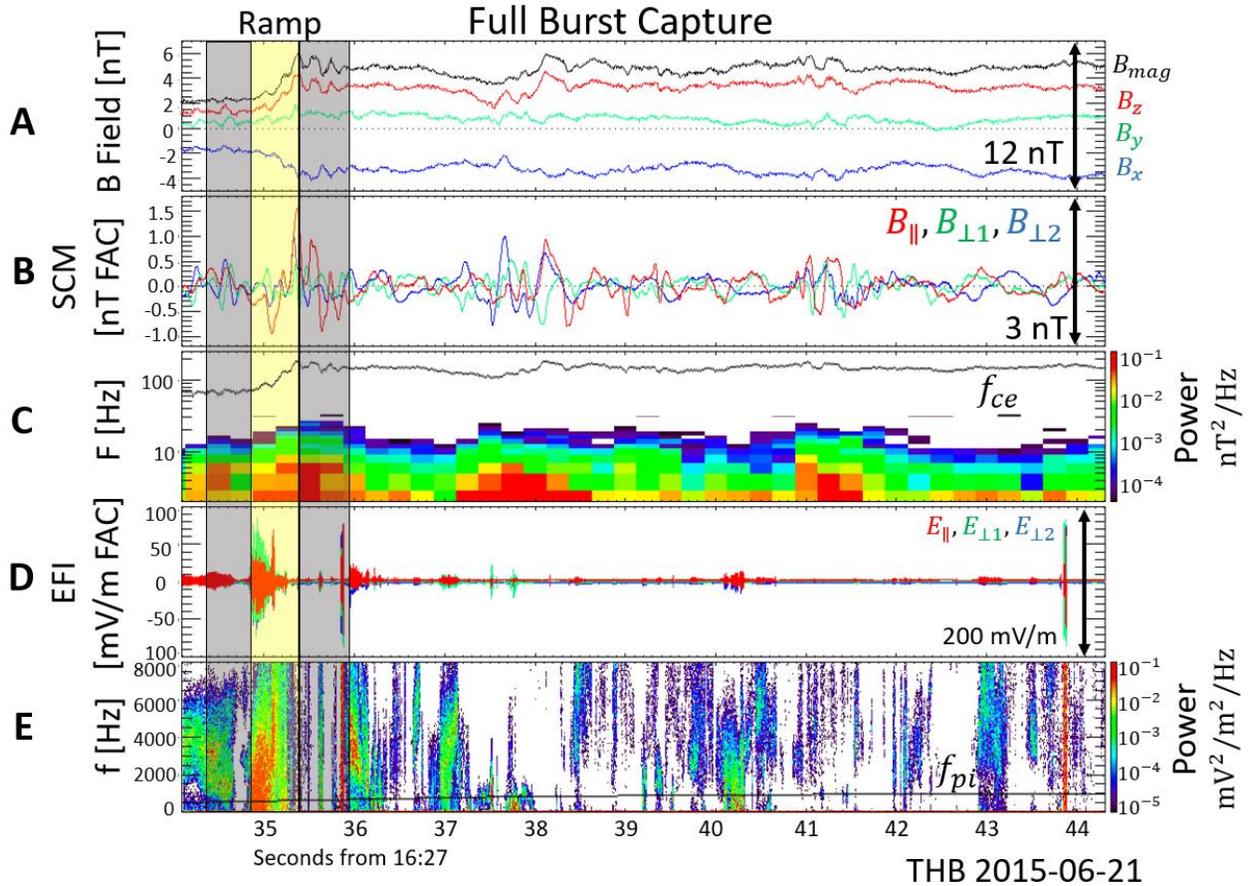

Figure 5: The 10 s wave burst capture that covers the entire transition region (grey) and ramp (yellow) for the 2015-06-21 shock event. (a) The background magnetic field in GSE coordinates. (b) The magnetic field waveform in FAC. (c) The total magnetic field power spectrum with the electron cyclotron frequency overlaid. (d) The electric field waveform in FAC. (e) The total electric field power spectrum with the ion plasma frequency overlaid.

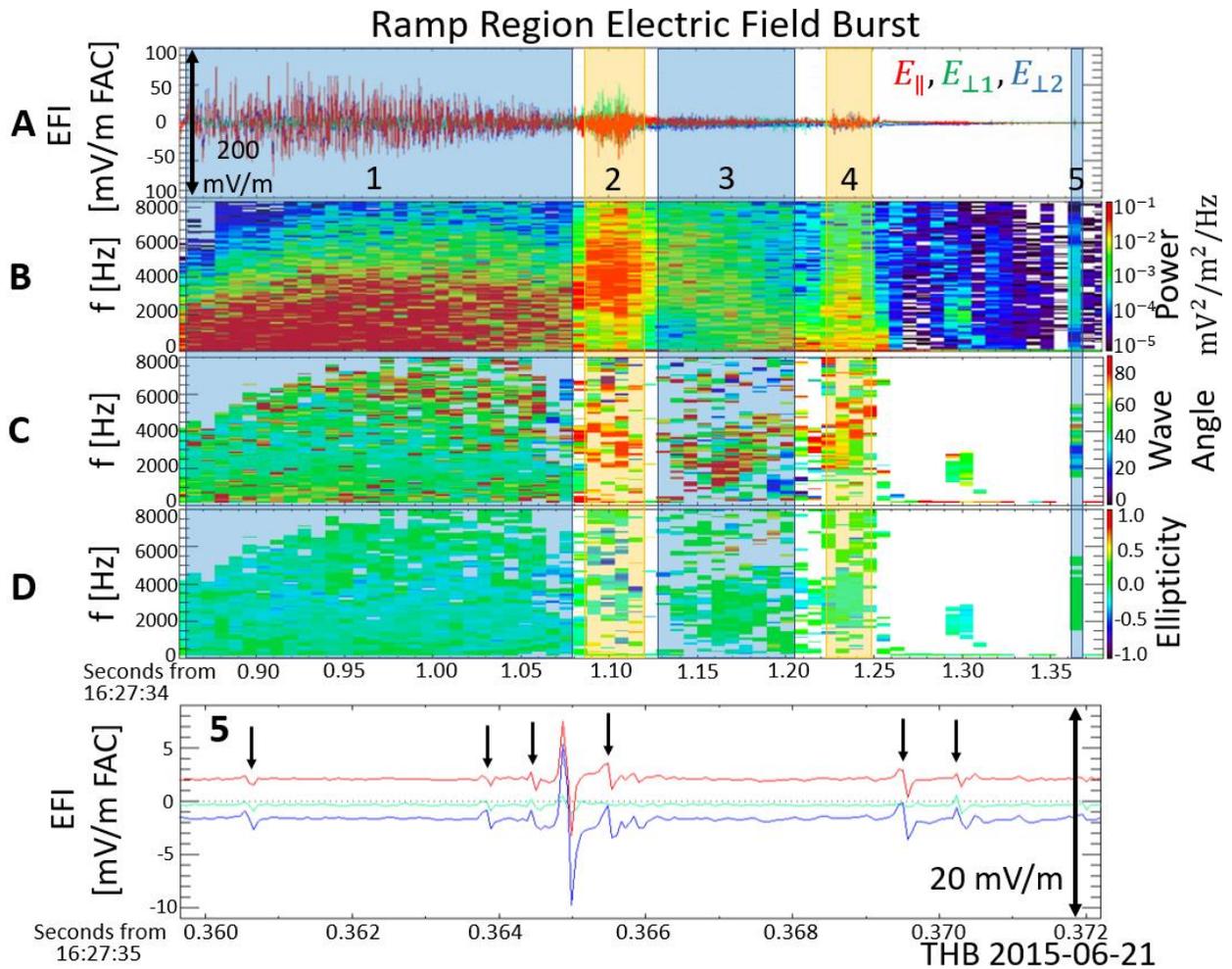

Figure 6: Zoom-in of the ramp region. Shaded regions 1-5 are different wave packets. (a) Electric field waveform in FAC. (b) Total electric field power spectrum. (c) Wave angle with respect to the magnetic field. (d) Ellipticity. (5) Zoom-in electric field burst of region 5, showing a solitary wave and nearby, smaller-amplitude solitary waves (black arrows).

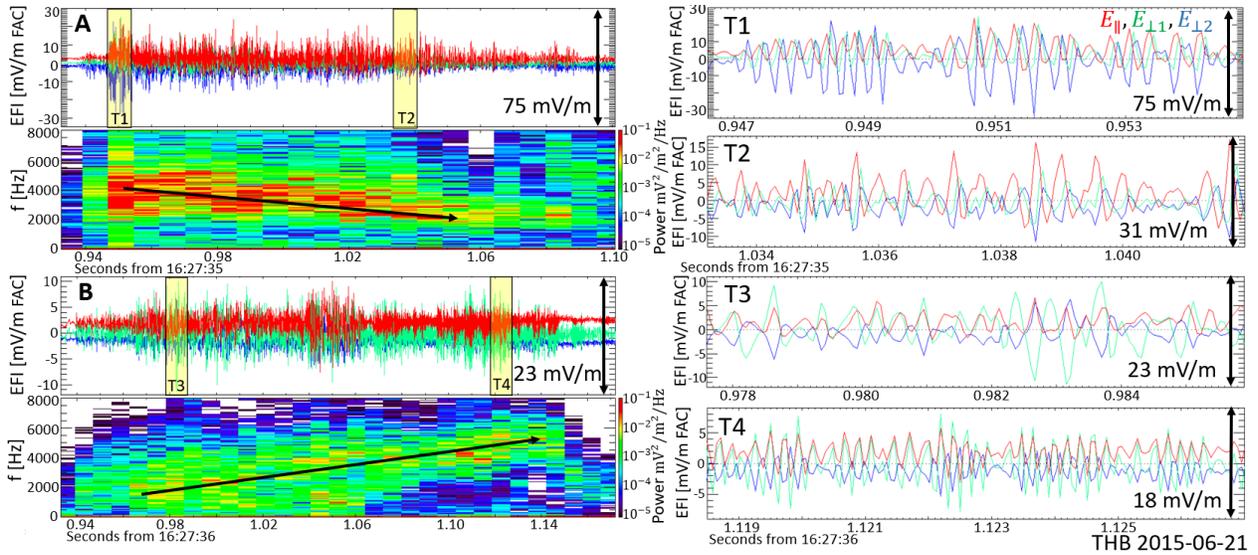

Figure 7: The electric field wave burst waveform data and power spectra for two TFES waves. (a) Decreasing frequencies with time can be seen in the power spectrum. (b) Increasing frequencies can be seen. (T1-T4) Electric field waveforms corresponding to the shaded times in (a) and (b).

### 3.2 Perturbed Shock

An overview of the 2017-05-20 shock observed by THB is shown in Figure 8 (same format as Figure 4). This event was a reverse shock with a ramp at roughly 09:52:47 (8d) and had a wave burst capture overlapping the transition region, beginning immediately upstream (<1 $\rho_{gi}$) of the ramp (8e). A clear enhancement was seen for electrons below 100 eV (8a). Upstream of the shock, electron pitch angle distributions (8b.1-3) show a population of electrons, likely strahl, traveling antiparallel, away from the sun, for energies between ~10-500 eV. Another population is shown travelling parallel, towards the sun, for energies between ~500-5000 eV. Downstream of the shock, there was a strong perpendicular enhancement of electrons for energies between ~100-3000 eV which lasted for roughly 8 minutes (~435 $\rho_{gi}$); this enhancement was absent for energies below 100 eV.

This event had a burst capture ~150 s upstream (~135 $\rho_{gi}$) of the ramp (9c) and one which covered part of the transition region, starting within 1 s upstream (<1 $\rho_{gi}$) of the ramp (9b). There were also two burst captures ~4 minutes (~217 $\rho_{gi}$) downstream, but they contained no waves of significant power and are not shown. Note that no burst covered the ~1 s duration of the ramp region itself. Figure 9 shows the magnetic field around the ramp region (shaded in yellow) and the transition region (shaded in grey) (9a). A perturbed shock with a clear whistler precursor is seen. The amplitude and frequency of this precursor were ~8 nT and ~1.6 Hz; the difference in the average upstream and downstream background magnetic fields was ~6.6 nT. Both burst captures and their respective power spectra are shown in Figure 9. This event had fewer and lower amplitude wave packets compared to most other events in this study. Only one wave packet with sufficient power was observed within the transition region, occurring ~2.5 seconds (~2 $\rho_{gi}$) upstream of the ramp. The color scale in the power spectra of Figure 9 is the same as in Figure 5, 6, and 7 to highlight the lack of large amplitude (>20 mV m$^{-1}$) waves. Seven wave modes were identified upstream, six ion acoustic waves and one solitary wave. All waveforms had amplitudes between ~2-8 mV m$^{-1}$. Note that the periodic spikes (some denoted by black arrows in 9c), which appear in both burst captures, were only observed by a single voltage probe, V5, indicating that the spikes are artificial. Similar to those shown in Figure 4e, the high power broadband spikes in both burst captures were also determined to be artificial. THC (~17000 km, ~85 $\rho_{gi}$ from THB), which observed a similar transition region, took two burst captures ~8 minutes (~980 $\rho_{gi}$) downstream, containing no waves above the power threshold of our study.

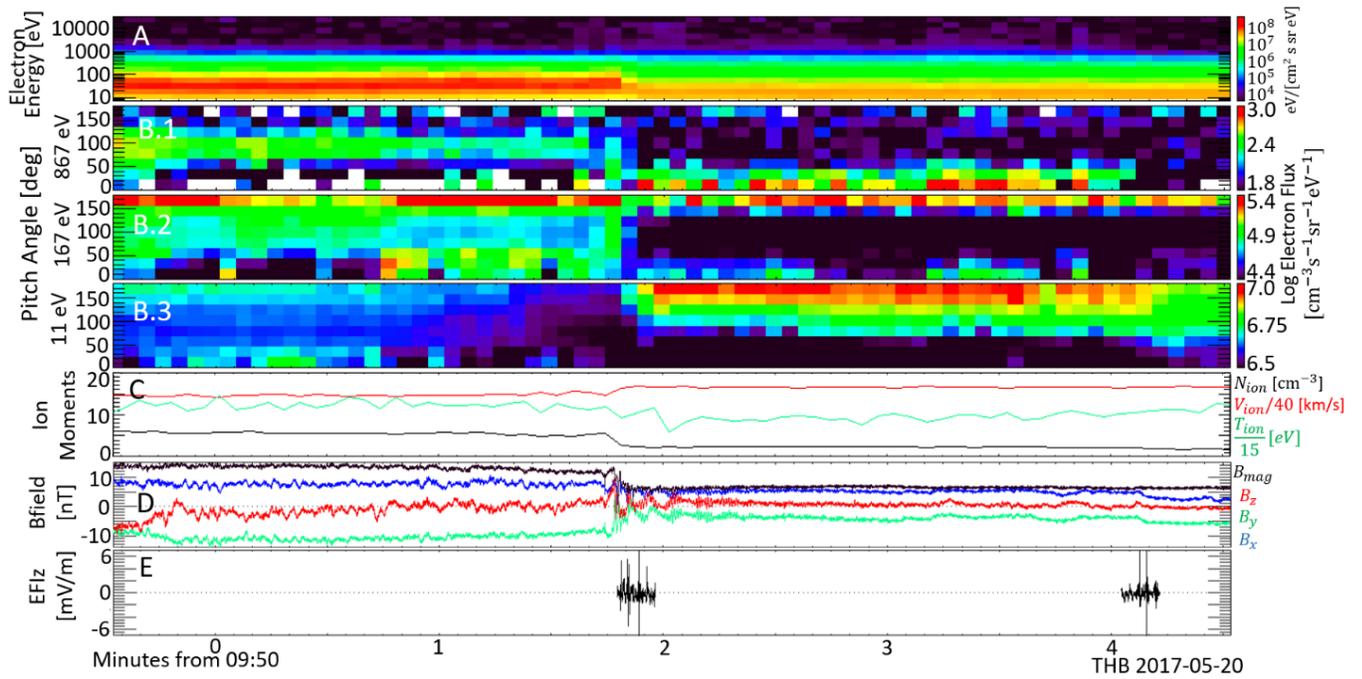

Figure 8: An overview of the 2017-05-20 shock event as observed by THB. (a) Electron energy flux from ~10 eV to 30 keV from ESA. (b.1-3) ESA electron pitch angle distributions for 11 eV, 167 eV, and 867 eV. Color bar is in units of log electron flux. (c) The ion density (cm⁻³, black), velocity (km s⁻¹, red, divided by 40), and temperature (eV, green, divided by 15). (d) Despun background magnetic field in GSE coordinates along with the magnitude (black). (e) Parallel component of the electric field wave burst captures.

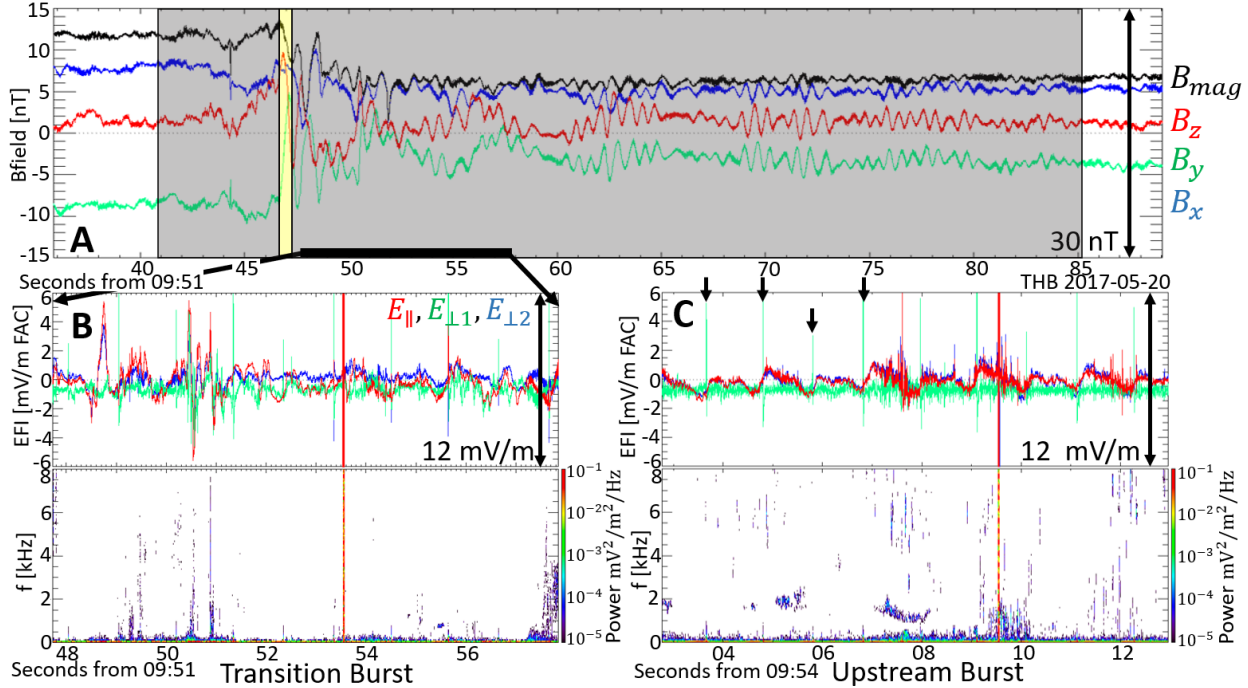

Figure 9: Two 10 s electric field wave burst captures near the ramp of the 2017-05-20 shock event. The ramp region is shaded yellow. The transition region is shaded in grey. (a) The background magnetic field. A whistler precursor is clearly evident. (b) The electric field waveform wave burst capture in field aligned coordinates (red is *z*, green is *y*, blue is *x*). This burst overlaps the transition region. (c) The second upstream electric field burst capture in field aligned coordinates. Arrows above the waveform data highlight the periodic spikes deemed artificial. Each burst capture waveform has its corresponding power spectrum shown below.

## 4 Statistical Results

The number of identified wave packets observed near laminar and perturbed shock ramps is summarized in Table 3. During the 13 events, there were ~290 s of burst capture time, corresponding to 26 individual burst captures within 10 minutes (average of 3330 $\rho_{gi}$ upstream, 1930 $\rho_{gi}$ downstream) of each ramp; 56% of the burst time occurred within 500 $\rho_{gi}$ of the ramp and 80% occurred within 1500 $\rho_{gi}$. To compare the occurrence rate of different wave modes between the four regions and two shock types, we normalized the total number of observed waves in each wave mode by the total burst capture time for a given region and shock type. Table 3 lists these rates for each shock type, shock region, and wave mode. These rates only include waves with sufficient power ($\gtrsim 10^{-3}$ mV$^2$ m$^{-2}$ Hz$^{-1}$ or an amplitude roughly $\gtrsim 2$ mV m$^{-1}$) to be identified by the algorithm. The total number of events with wave burst captures for each shock type and region is shown in parentheses in the most left column. In the four central columns, the total number of wave mode identifications made for a given shock type, region, and wave mode category is listed in parentheses next to the occurrence rate. The last column lists the total burst capture time for a given region and shock type. Most of the burst capture time and wave packets occurred downstream of both shock types, with both types having a similar amount of burst capture time.

The major difference between the two shock types is in the occurrence rate of the different wave modes. In the downstream region of perturbed shocks, the occurrence rate for ion acoustic waves was ~2.7 times that for laminar shocks; TFES waves had ~2.3 times the occurrence rate. For solitary waves and whistler modes, the statistics are less significant due to the lower number of observations. The occurrence rate for solitary waves was ~7 times greater downstream of perturbed shocks compared to laminar shocks; for whistler waves, the occurrence rate was ~10 times greater.

For six events (three laminar and three perturbed), burst captures occurred during the transition region. Both shock types had similar burst capture times in this region, and the occurrence rate for ion acoustic and whistler mode waves was similar for both types. For TFES waves, however, the occurrence rate within the transition region for perturbed shocks was nearly triple that for laminar shocks. This difference may provide a clue to the physical mechanism that results in the rapid frequency change. The average (median) amplitude for waves in the transition region for laminar shocks was ~22 (~12) mV m$^{-1}$; for perturbed shocks, it was ~16 (~10) mV m$^{-1}$. Note that one perturbed shock with burst data in the transition region did not observe any waves with power above $10^{-3}$ mV$^2$ m$^{-2}$ Hz$^{-1}$. In the transition region, waves near laminar shocks on average have higher amplitudes than waves near perturbed shocks.

Three events (two laminar and one perturbed) had burst captures covering the entire ramp region. For these two laminar shocks, the occurrence rate of waves in the ramp region was significantly higher than either up- or downstream. The peak amplitudes for both laminar shocks were due to ion acoustic waves, and the amplitudes were, respectively, ~169 mV m$^{-1}$ and ~9 mV m$^{-1}$. For the perturbed event, no wave with amplitude $\gtrsim 2$ mV m$^{-1}$ was observed within the ramp region. The low number of events preclude a definitive conclusion about waves within the ramp regions of perturbed shocks.

| Shock Type, Region | Ion Acoustic | TFES | Solitary Waves | High $f$ Whistlers | Burst Time |
|---|---|---|---|---|---|
| Laminar, Upstream (1) | - (0) | 3.333 (1) | - (0) | - (0) | 0.3 s |
| Laminar, Transition (3) | 0.612 (20) | 0.183 (6) | 0.153 (5) | 0.061 (2) | 32.7 s |
| Laminar, Ramp (2) | 5.000 (6) | 1.667 (2) | 3.333 (4) | - (0) | 1.2 s |
| Laminar, Downstream (5) | 0.328 (36) | 0.164 (18) | 0.018 (2) | 0.018 (2) | 109.8 s |
| Perturbed, Upstream (1) | 0.588 (6) | - (0) | 0.098 (1) | - (0) | 10.2 s |
| Perturbed, Transition (3) | 0.591 (22) | 0.511 (19) | 0.081 (3) | 0.054 (2) | 37.2 s |
| Perturbed, Ramp (1) | - (0) | - (0) | - (0) | - (0) | 1.3 s |
| Perturbed, Downstream (7) | 0.884 (90) | 0.373 (38) | 0.118 (12) | 0.079 (8) | 101.8 s |
| Total | 174 | 82 | 23 | 14 | 292.0 s |

Table 3: Occurrence rate (in units of number of waves sec$^{-1}$) for a given shock type, region, and wave mode. The total number of waves, shown in parentheses next to the rate, was normalized by the total burst capture length for each region. In the left column, the number of events with observations for each shock type and region is shown in parentheses.

The number of wave packets observed was not equally distributed among the events in our dataset. Three events, one laminar (THB 2015-06-21) and two perturbed (THC 2015-06-24; THC 2015-06-27), observed significantly more wave packets than other events. When these events are removed, laminar shocks have higher occurrence rates than perturbed shocks for all regions that have burst capture data to compare. No strong dependence of wave occurrence rates based on $M_f$ or $\theta_{Bn}$ was observed. Any observed differences were more likely due to the limited number of events than any relationship to either $M_f$ or $\theta_{Bn}$. Significant numbers of additional events, beyond the scope of this study, would be required to provide statistically significant conclusions on $\theta_{Bn}$ or $M_f$ dependence.

The high frequency whistler mode waves observed (~0.5-3 mV m$^{-1}$; ~50-1000 pT) had smaller electric field amplitudes compared to other wave modes in this study, but also had some of the longest durations (~0.1-10 s; ~8-2000 wave periods). The magnetic field amplitudes are similar to, but in some cases higher, than the range of amplitudes (~10-100 pT, $\delta B/B$ ~0.1-1%) found in other studies of high frequency whistlers in the solar wind [*Lacombe et al.*, 2014; *Tong et al.*, 2019]. High frequency whistlers were seen in only three events (2013-07-09, 2015-06-27, and 2017-02-24). Even though most of the observed whistlers were ≲ 2 mV m$^{-1}$, they were included in the statistics because they were observed in an identified wave packet, whether with another wave mode or individually. In addition, the magnetic amplitudes $\delta B$ were as large as 7% of the background field $B$. It is possible that other events had whistlers, but either those whistlers did not have high enough amplitude to trigger a wave packet or did not occur simultaneously with another higher amplitude wave mode. One whistler, with an amplitude of ~0.5 mV m$^{-1}$ and ~0.15 nT ($\delta B/B$ ~0.6%) and a frequency ~200 Hz (0.43 $f_{ce}$), was observed ~1 minute (~320 $\rho_{gi}$) downstream on 2013-07-09 with a duration at least as long as the 10-second wave burst capture. A second whistler, with an amplitude of ~0.5 mV m$^{-1}$ and ~0.1 nT ($\delta B/B$ ~1%) and a frequency ~160 Hz (0.36 $f_{ce}$), occurred ~3 minutes (~960 $\rho_{gi}$) downstream and also lasted the full duration of the 10-second wave burst capture. Both of these long duration whistlers had a ~10° wave angle and had right-handed (~0.7) ellipticity, consistent with *Lacombe et al.* [2014].

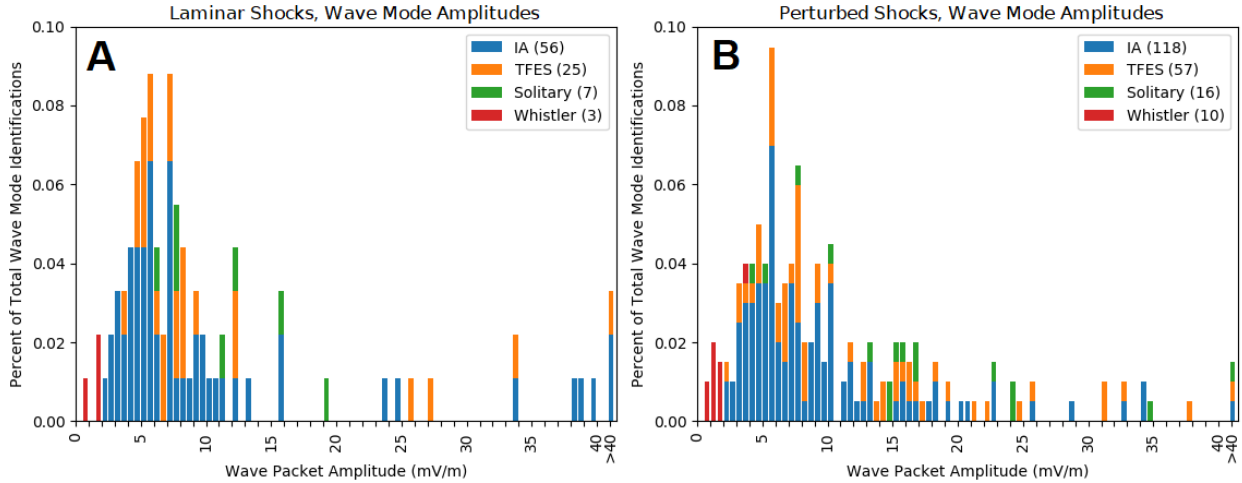

Figure 10: The amplitude distributions of wave packets near (a) laminar shocks and (b) perturbed shocks. Wave packets containing ion acoustic (blue), TFES (orange), solitary (green), and whistler mode (red) waves are shown separately in each distribution.

Figure 10 shows the distribution of the amplitudes of individual wave packets for laminar and perturbed shocks, subdivided into ion acoustic (blue), TFES (orange), solitary (green), and high frequency whistler mode (red) waves. Most wave packets (193/279 ≈ 69%) had amplitudes between 2 and 10 mV m$^{-1}$. The average (median) amplitude of the wave modes observed was 10.2 (6.6) mV m$^{-1}$ for ion acoustic, 12.0 (7.5) mV m$^{-1}$ for TFES, 15.3 (14.4) mV m$^{-1}$ for solitary waves, and 1.1 (0.9) mV m$^{-1}$ and 0.42 (0.32) nT for whistler modes. No significant difference in amplitude was found between laminar and perturbed shocks. Note, however, that large amplitude waves were more concentrated near the ramp region for laminar shocks, though the limited number of shocks with burst data covering and adjacent to the ramp limit the possible significance of this observation. Similarly, no significant dependence of amplitude on either $M_f$ or $\theta_{Bn}$ was found.

Differences in the duration of ion acoustic and TFES wave modes between shock types were examined using the duration of wave packets normalized by the average upstream ion plasma frequency. The resulting distribution for both wave modes peaked around 5-10 average upstream ion plasma periods with a long tail towards longer durations. There was no difference between laminar and perturbed shocks. For all wave packets analyzed, including whistler and solitary waves, most packets (247/279 ≈ 89%) lasted for less than 100 ms; the median duration was ~31 ms. These short durations indicate that using filter bank data for identifying wave modes in ramps may be problematic. Typically, the highest sampling filter bank data has a ~1/8 or 1/16 second timescale (roughly 125 or 63 ms) [*Cully et al.*, 2008; *Wygant et al.*, 2013]. The short duration of a large portion of observed packets (223/279 ≈ 80% had durations less than 60 ms) provides evidence that averaged filter bank data cannot be used to accurately assess wave modes in or near shocks.

## 5 Discussion

Three shocks (two laminar, one perturbed) had burst captures that overlapped the shock ramp. In the ramps of the two laminar shocks, ion acoustic, TFES, and solitary waves were observed. The occurrence rates for these wave modes were 10-20 times greater in the ramp region than downstream. Although a small number of shocks were analyzed, the high occurrence rate of waves in both of the observed laminar shock ramps, and not in the perturbed shock ramp, provides support for the idea that energy dissipation through wave-particle interactions is more prominent in laminar shocks than perturbed shocks. The maximum amplitude waves for the two laminar shocks with waves within the ramp were ion acoustic modes and were ~169 mV m$^{-1}$ and ~9 mV m$^{-1}$, respectively. Note that these large amplitudes were observed within single wave packets. To assess the role of wave-particle interactions for these two events, we took the ratio of the energy dissipation rates due to wave-particle interactions (the cumulative sum of -j$_0$·δE across the shock ramp, where $j_0$ is the current density and $\delta E$ is the fluctuations in the electric field) to the change in kinetic energy density across the shock ramp ($\frac{1}{2}\frac{\Delta(\rho U_{shn}^2)}{\Delta\tau}$, where $\rho$ is the scalar mass density, $U_{shn}$ is the shock normal speed in the plasma rest frame, and $\Delta\tau$ is the shock ramp duration in the Normal Incidence Frame); this is the same ratio as Equation 3b in *Wilson III et al.* [2014a]. For the two laminar shocks discussed above, this ratio was ~567 and ~5.3, respectively. The fact that the ratio of energy dissipation to the change in kinetic energy is greater than one provides evidence that these laminar shocks could, depending on the efficiency of the energy/momentum exchange, dissipate energy solely through wave-particle interactions. In comparison, the single perturbed shock with burst data in the ramp had a ratio of ~1.0 and no waves with amplitudes ≳ 2 mV m$^{-1}$ were observed in the ramp region. Although it is possible that the single perturbed shock may have dissipated the required energy through wave-particle interactions, this dissipation process appears to play a more prominent role in the laminar shocks.

Downstream of shocks, where most of the burst capture data was obtained, there is a stark difference in the wave occurrence rate between laminar and perturbed shocks. When normalized by the amount of burst capture time, perturbed shocks were observed with significantly more wave packets downstream than laminar shocks, with the occurrence rate of ion acoustic and TFES waves downstream of perturbed shocks being ~2-3 times that downstream of laminar shocks. Both solitary waves and whistler mode waves were also much more common. These higher occurrence rates are due to two events with significant wave packet generation. The high occurrence rate may be due to increased ion heating, as seen in our perturbed events, and in the study by *Wilson III et al.* [2012], who also reported gyrating ions downstream of a perturbed shock which could provide free energy for the waves. Further analysis of ion/electron heating or the determination of free energy sources is beyond the scope of this study. Future investigation of sources of free energy near these structures could provide further insight into the difference in occurrence rates. For the other perturbed events, the wave occurrence rates are low in all regions, suggesting that the source of free energy for the two perturbed events with high occurrence rates is not present at all perturbed shocks.

No significant difference was found in either the wave amplitudes or wave packet durations between the two shock classifications. Similarly, no amplitude or duration dependence on $\theta_{Bn}$ or $M_f$ was seen. The average (median) duration of all wave packets was 46 ms (23 ms). A large portion (80%) of wave packets had a duration less than 60 ms. This short duration provides evidence that average filter bank data cannot be used to accurately identify wave modes in or near shocks. Initial examination of ion distributions showed features consistent with reflected ion beams for a portion of shocks in this study; these features were present for an equal number of laminar and perturbed shocks. Further analysis of ion distributions was beyond the scope of this study. Although the small number of events in this study precludes a definite result that applies to all low Mach number, quasi-perpendicular IP shocks, it is sufficient to study the differences in wave activity in the range of ~50-8000 Hz between laminar and perturbed shocks.

Recent findings at the high Mach number ($M_f$ ~7), terrestrial bow shock by *Goodrich et al.* [2018] using MMS observations found similar wave amplitudes and durations to those in this study. *Goodrich et al.* [2018] observed a number of ion acoustic waves in the ramp of an oblique bow shock crossing, with amplitudes occasionally exceeding 100 mV m$^{-1}$ and durations between 10-100 ms. This is consistent with our findings of ion acoustic mode amplitudes and durations near laminar shocks. In addition to the waves observed in the ramp region, observed reflected ions, which they suggest may provide free energy, may also play a role in energy dissipation for this high Mach number bow shock crossing. *Goodrich et al.* [2018] also observed solitary waves with amplitudes between 100-300 mV m$^{-1}$, much greater than the amplitudes observed near shocks in this study. The variability of wave modes observed in the ramp region of the bow shock by *Goodrich et al.* [2018], which includes whistler mode waves, solitary waves, ion acoustic waves, and Bernstein mode waves potentially generated by the EDCI, is similar to the variability of wave modes seen near the lower Mach number IP shocks in this study.

## 6 Conclusions

Eleven quasi-perpendicular interplanetary shocks observed by the ARTEMIS spacecraft with electric and magnetic wave burst captures within 10 minutes (average of ~3330 $\rho_{gi}$ upstream and ~1930 $\rho_{gi}$ downstream, depending on local plasma parameters), either upstream or downstream, of the ramp region were analyzed. Two shocks had burst data observations at both spacecraft, yielding a total of 13 events. The shock structure was classified as either laminar or perturbed. Perturbed shocks were characterized by whistler precursors with an amplitude $\delta B$ on the order of the difference between the upstream and downstream average magnetic field magnitudes $\Delta B$ ($\delta B/\Delta B \gtrsim 1/3$). This is the first study to examine the differences in waves from ~50-8000 Hz for laminar and perturbed shocks.

The burst captures taken near or in the ramp region of these shocks enabled a study of waves in the frequency range of ~50-8000 Hz for the electric field and ~1-500 Hz for the magnetic field; waves with power (amplitude) above $10^{-3}$ mV$^2$ m$^{-2}$ Hz$^{-1}$ ( $\gtrsim$ 2 mV m$^{-1}$) were analyzed. The wave modes observed included ion acoustic waves, TFES waves, solitary waves, and high frequency (~0.1-0.4 $f_{ce}$) whistler mode waves. From this study, we conclude:

1. A wide range of wave modes are observed, including ion acoustic, TFES, solitary, and whistler waves, for both laminar and perturbed shocks. This is most apparent in the ramps of the two laminar shocks with burst data covering the ramp region, where there were high wave occurrence rates.

2. In the downstream region, two perturbed shocks had 2-3 times the occurrence rate of waves than laminar shocks. Other perturbed events lacked waves of significant power, suggesting that the source of free energy for the two perturbed shocks with high occurrence rates is not present for all perturbed shocks.

3. TFES waves (waves with characteristics similar to ion acoustic waves, but with quick changes in frequency with time) were observed ~2-3 times as often in the transition region and downstream of perturbed shocks than laminar shocks. This may provide a clue to the physical mechanism that results in the frequency change.

4. No significant differences in average wave amplitude or wave packet duration for waves with frequencies between 50 and 8000 Hz was found between laminar and perturbed shocks.

We have shown that laminar shocks are observed with a wide variety of large amplitude wave modes within the transition region and near the ramp region. Further, the observed wave occurrence rates are higher in the transition region than downstream for laminar shocks. The wave occurrence rates downstream of perturbed shocks are at least twice as much as downstream of laminar shocks. The difference in the region of highest wave occurrence suggests that the primary mechanism of energy dissipation may differ for laminar and perturbed shocks. Further research utilizing a larger number of shocks with burst data covering the ramp region and investigating particle and field effects will provide more insight into the differences between laminar and perturbed shocks.

**Acknowledgements**

The ARTEMIS data is publicly available from CDAWeb, as well as other sources through the use of the SPEDAS software [*Angelopoulos et al.*, 2019]. This research was supported by NASA grants NNX16AF80G, NNX14AK73G, and 80NSSC19K0305. L.B.W. was partially supported by *Wind* MO&DA grants and a Heliophysics Innovation Fund (HIF) grant. The work was also supported by the International Space Science Institute's (ISSI) International Teams programme.

**Appendix A: Wave Packet ID Algorithm**

In any given burst capture, 0 to ≳100 wave packets meeting the power criterion could be found. We implemented an algorithm to guarantee that each wave packet was identified with a consistent set of criteria, namely a power threshold in the electric field. The algorithm used the total power in the electric field and marked time-frequency pairs (each has dimensions of ~8 ms by ~66 Hz) when the power exceeded a preset threshold. For this study, the threshold was set to $10^{-3}$ mV$^2$ m$^{-2}$ Hz$^{-1}$; typical background values were $\lesssim 10^{-5}$ mV$^2$ m$^{-2}$ Hz$^{-1}$. The boundary of a wave packet, which is made up of grouped, adjacent marked pairs, occurred where there were no longer marked pairs adjacent in either the time or frequency domain. The boundary in the time domain provided the time interval of the wave packet. The algorithm can identify waveforms with different characteristic frequencies in the same time interval, allowing identification of multiple wave modes. Low amplitude packets ($\lesssim$ 2 mV m$^{-1}$) and/or packets with broadband power, such as solitary waves, are not well identified by the algorithm and are excluded from this study.

Figure 11 shows the steps in this algorithm. The first panel shows the electric field wave burst data (11a), and the end product of the identification of wave packets, each shown by a black box. The second panel shows the power spectrum of wave burst data (11b), with red ovals showing the rough boundaries in time and frequency of identified wave packets.

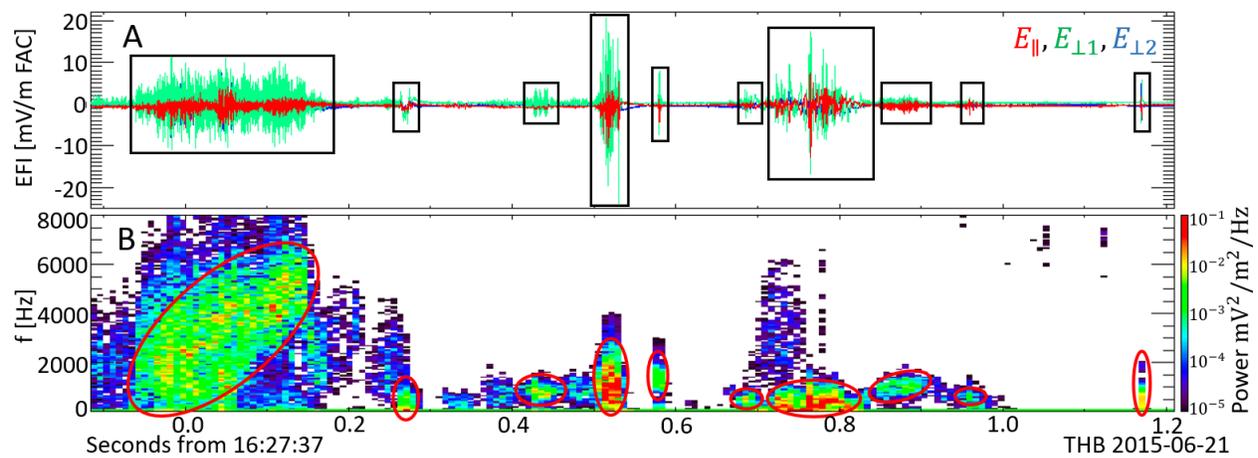

Figure 11: (a) Electric field data from a wave burst capture. Waveforms inside of black boxes are what were considered wave packets. (b) The total electric field burst power spectrum. Red ovals indicate the rough time and frequency boundaries of each wave packet.

Each initial wave packet was then fit with a convex hull to find its peak-to-peak amplitude. For the upper (lower) bound, the maximum (minimum) value in a 1 ms interval was used to fit the bounds. The amplitude was then found by taking the difference between the upper and lower bounds. Figure 12 shows an example of the convex hull surrounding a wave packet. The maximum amplitude and duration of each wave packet, along with the number of time-frequency pairs, was output to a list at the end of the algorithm.

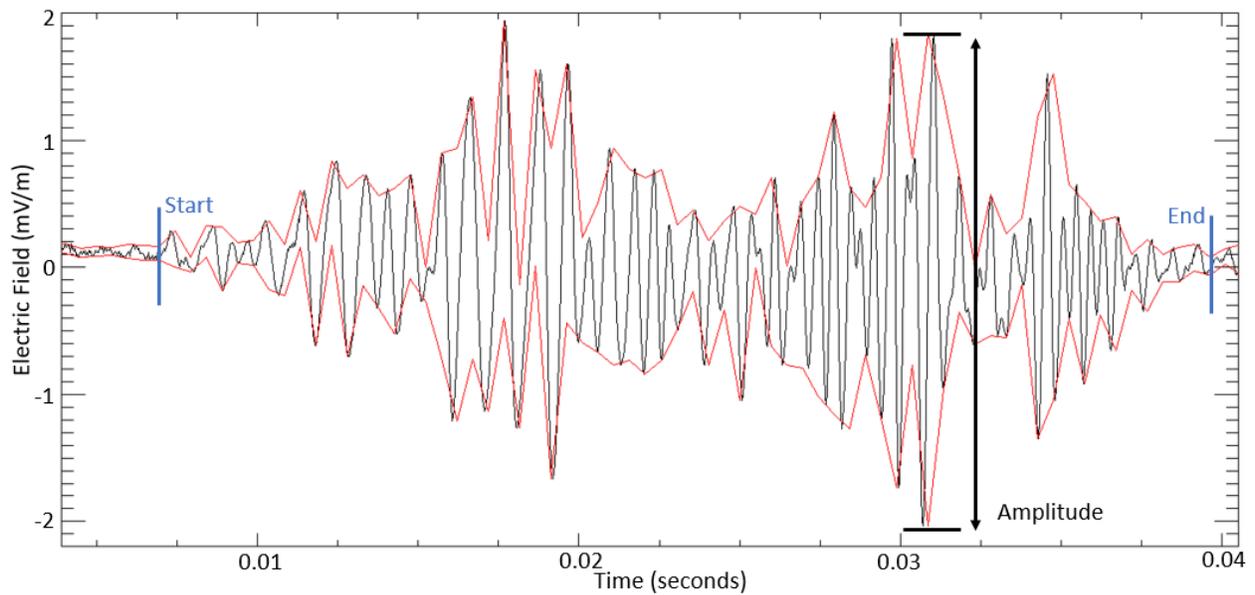

Figure 12: The convex hull (red) fitting for a wave packet. The maximum amplitude (black arrow) and starting and ending points (blue lines) of the wave packet are labeled.

Waveforms with power near the threshold, but not consistently above it, would often generate individual or small clumps (<5) of time-frequency pairs. Waveforms with borderline power were removed from the initial list of candidate wave packets by removing wave packets with <5 pairs. The remainder were then carefully sorted such that no two wave packets contained the same waveform in order to avoid double counting. Lastly, clear examples of artificial signals, such as features only observed by a single boom pair in the raw field data, were removed. This final list was used in this study.

**References**

Abraham-Shrauner, B., and S. H. Yun (1976), Interplanetary Shocks Seen by AMES Plasma Probe on Pioneer 6 and 7, *J. Geophys. Res.*, 81, 2097–2102, doi:10.1029/JA081i013p02097

Akimoto, K. and D. Winske (1985), Ion-Acoustic-Like Waves Excited by the Reflected Ions at the Earth's Bow Shock, *J. Geophys. Res.*, 90(A12), 12,095–12,103, doi:10.1029/JA090iA12p12095

Andersson, L., et al. (2009), New Features of Electron Phase Space Holes Observed by the THEMIS Mission, *Phys. Rev. Lett.*, 102(22), 225004, doi:10.1103/PhysRevLett.102.225004

Angelopoulos, V. (2011), The ARTEMIS Mission, *Space Sci. Rev.*, 165, 3-25, doi:10.1007/s11214-010-9687-2

Angelopoulos, V. et al. (2019), The Space Physics Environment Data Analysis System (SPEDAS), *Space Sci. Rev.*, 215:9, doi:10.1007/s11214-018-0576-4

Artemyev, A. V., V. Angelopoulos, and J. M. McTiernan (2018), Near-Earth Solar Wind: Plasma Characteristics From ARTEMIS Measurements, *J. Geophys. Res. Space. Phys.*, 123, 9955-9962, doi:10.1029/2018JA025904

Auster, H. U., et al. (2008), The THEMIS Fluxgate Magnetometer, *Space Sci. Rev.*, 141, 235-264, doi:10.1007/s11214-008-9365-9

Bale, S. D., et al. (1998), Bipolar electrostatic structures in the shock transition region: Evidence of electron phase space holes, *Geophys. Res. Lett.*, 25, 2929-2932, doi:10.1029/98GL02111

Bale, S. D., et al. (2005), Quasi-perpendicular shock structure and processes, *Space Sci. Rev.*, 118, 161-203, doi:10.1007/s11214-005-3827-0

Balikhin, M., S. Walker, R. Treumann, H. Alleyne, V. Krasnoselskikh, M. Gedalin, M. Andre, M.
Dunlop, and A. Fazakerley (2005), Ion sound wave packets at the quasiperpendicular shock front, *Geophys. Res. Lett.*, 32, L24106, doi:10.1029/2005GL024660

Beinroth, H. J. and F. M. Neubauer (1981), Properties of Whistler Mode Waves Between 0.3 and 1.0 AU from Helios Observations, *J. Geophys. Res.*, 86, 7755-7760, doi:JA086iA09p07755

Biskamp, D. (1973), Collisionless shock waves in plasmas, *Nucl. Fusion*, 13, 719-740, doi:10.1088/0029-5515/13/5/010

Bonnell, J. W., et al. (2008), The Electric Field Instrument (EFI) for THEMIS, *Space Sci. Rev.*, 141, 303-341, doi:10.1007/s11214-008-9469-2

Breneman, A., C. Cattell, S. Schreiner, K. Kersten, L.B. Wilson III, P. Kellogg, K. Goetz, and L. K. Jian (2010), Observations of large-amplitude, narrowband whistlers at stream interaction regions, *J. Geophys. Res.*, 115, A08104, doi:10.1029/2009JA014920

Breneman, A., C. Cattell, K. Kersten, A. Paradise, S. Schreiner, P. J. Kellogg, K. Goetz, and L. B. Wilson III (2013), STEREO and Wind observations of intense cyclotron harmonic waves at the Earth's bow shock and inside the magnetosheath, *J. Geophys. Res. Space Physics*, 118, 7654-7664, doi:10.1002/2013JA019372

Cattell, C., J. Crumley, J. Dombeck, J. Wygant, and F. S. Mozer (2002), Polar observations of solitary waves at the Earth's magnetopause, *Geophys. Res. Lett.*, 29(5), 1065, doi:10.1029/2001GL014046

Cattell, C., et al. (2005), Cluster observations of electron holes in association with magnetotail reconnection and comparison to simulations, *J. Geophys. Res.*, 110, A01211, doi:10.1029/2004JA010519

Cattell, C. A., B. Short, A. W. Breneman, and P. Grul (2020), Narrowband Large Amplitude Whistler-mode Waves in the Solar Wind and Their Association with Electrons: STEREO Waveform Capture Observations, *ApJ*, 897, 126, doi:10.3847/1538-4357/ab961f

Chen, L.-J., et al. (2018), Electron Bulk Acceleration and Thermalization at Earth's Quasiperpendicular Bow Shock, *Phys. Rev. Lett., 120(22)*. doi:10.1103/physrevlett.120.225101

Cohen, I. J., et al. (2019), High-Resolution Measurements of the Cross-Shock Potential, Ion Reflection, and Electron Heating at an Interplanetary Shock by MMS, *J. Geophys. Res. Space Physics*, doi:10.1029/2018JA026197

Cohen, Z. A., C. A. Cattell, A. W. Breneman, L. A. Davis, P. Grul, K. Kersten, L. B. Wilson III, and J. R. Wygant (2020), The Rapid Variability of

Waves in Interplanetary Shocks: STEREO Observations, *Astrophys. J.*, 904,174, doi: 10.3847/1538-4357/abbeec

Coroniti, F. V. (1970a), Dissipation discontinuities in hydromagnetic shock waves, *J. Plasma Phys.*, 4,265- 282, doi:10.1017/S0022377800004992

Coroniti, F. V. (1970b), Turbulence structure of high-beta perpendicular fast shocks, *J. Geophys. Res.*, 75, 7007- 7017, doi:10.1029/JA075i034p07007

Coroniti, F. V., C. F. Kennel, F. L. Scarf, and E. J. Smith (1982), Whistler Mode Turbulence in the Disturbed Solar Wind, *J. Geophys. Res.*, 87, 6029-6044, doi: 10.1029/JA087iA08p06029

Cully, C. M., R. E. Ergun, K. Stevens, A. Nammari, and J. Westfall (2008), The THEMIS Digital Fields Board, *Space Sci. Rev.*, 141(1-4), 343- 355, doi:10.1007/s11214-008-9417-1

Edmiston, J. P. and C. F. Kennel (1984), A parametric survey of the first critical Mach number for a fast MHD shock, *J. Plasma Phys.*, 32, 429-441, doi:10.1017/S002237780000218X

Ergun, R. E., et al. (1998), FAST satellite observations of large-amplitude solitary structures, *Geophys. Res. Lett.*, 25, 2041-2044, doi:10.1029/98GL00636

Fairfield, D. H. (1974), Whistler Waves Observed Upstream From Collisionless Shocks, *J. Geophys. Res.*, 79, 1368-1378, doi:10.1029/JA079i010p01368

Farris, M. H., C. T. Russell, and M. F. Thomsen (1993), Magnetic structure of the low beta, quasi-perpendicular shock, *J. Geophys. Res.*, 98(15), 15,285- 15,294, doi:10.1029/93JA00958

Filbert, P. C. and P. J. Kellogg (1979), Electrostatic noise at the plasma frequency beyond the Earth's bow shock, *J. Geophys. Res.*, 84(A4), doi: doi:10.1029/JA084iA04p01369

Formisano, V., Russell, C. T., Means, J. D., Greenstadt, E. W., Scarf, F. L., and Neugebauter, M. (1975), Collisionless shock waves in space: A very high beta structure, *J. Geophys. Res.*, 80, 2013- 2022, doi:10.1029/JA080i016p02013

Formisano, V., and P. C. Hedgecock (1973a), Solar wind interaction with the Earth's magnetic field: 3. On the Earth's bow shock structure, *J. Geophys. Res.*, 78, 3745-3760, doi:10.1029/JA078i019p03745

Formisano, V., and P. C. Hedgecock (1973b), On the structure of the turbulent bow shock, *J. Geophys. Res.*, 78, 6522-6534, doi:10.1029/JA078i028p06522

Formisano, V. and R. Torbert (1982), Ion Acoustic Wave Forms Generated by Ion-Ion Streams at the Earth's Bow Shock, *Geophys. Res. Lett.*, 9, 207-210, doi:10.1029/GL009i003p00207

Franz, J. R., P. M. Kintner, S. J. Pickett and L.-J. Chen (2005), Properties of small-amplitude electron phase-space holes observed by Polar, *J. Geophys. Res.: Space Physics*, 110(A9), doi:10.1029/2005ja011095

Fredricks, R. W., C. F. Kennel, F. L. Scarf, G. M. Crook, and I. M. Green (1968), Detection of electric-field turbulence in the Earth's bow shock, *Phys. Rev. Lett.*, 21(26), 1761-1764, doi:10.1103/PhysRevLett.21.1761

Fuselier, S. and D. A. Gurnett (1984), Short Wavelength Ion Waves Upstream of the Earth's Bow Shock, *J. Geophys. Res.*, 89, 91-103, doi:10.1029/JA089iA01p00091

Galeev, A. A. and V. I. Karpman (1963), Turbulence theory of a weakly nonequilibrium low-density plasma and structure of shock waves, *Sov. Phys. JETP*, 17(2), 403-409.

Gary, S. P. (1981), Microinstabilities upstream of the Earth's bow shock: A brief review, *J. Geophys. Res.*, 86, 4331-4336, doi:10.1029/JA086iA06p04331

Gary, S. P. and M. M. Mellott (1985), Whistler Damping at Oblique Propagation: Laminar Shock Precursor, *J. Geophys. Res.*, 90 99-104, doi:10.1029/JA090iA01p00099

Gary, S. P., W. C. Feldman, D. W. Forslund, and M. D. Montgomery (1975), Heat Flux Instabilities in the Solar Wind, *J. Geophys. Res.*, 80, 4197-4203. doi:10.1029/JA080i031p04197

Giagkiozis, I., S.N. Walker, and M.A. Balikhin (2011), Dynamics of ion sound waves in the front of the terrestrial bow shock, *Ann. Geophys.*, 29, 805- 811, doi:10.5194/angeo-29-805-2011

Giagkiozis, S. et al. (2018), Statistical study of the properties of magnetosheath lion roars, *J. Geophys. Res. Space Physics*, 123(7), 5435-5451, doi:10.1029/2018ja025343

Goncharov, O., J. Šafránková, Z. Němeček, L. Přech, A. Pitňa, and G. N. Zastenker (2014), Upstream and downstream wave packets associated with low-Mach number interplanetary shocks, *Geophys. Res. Lett.*, 41, 8100-8106, doi:10.1002/2014GL062149

Goodrich, K. A. et al. (2018). MMS observations of electrostatic waves in an oblique shock crossing. *J. Geophys. Res. Space Physics*, 123, 9430- 9442. doi:10.1029/2018JA025830

Goodrich, K. A., R. Ergun, S. J. Schwartz, L. B. Wilson III, A. Johlander, D. Newman, F. D. Wilder, J. Holmes, J. Burch, R. Torbert, Y. Khotyaintsev, P.-A. Lindqvist, R. Strangeway, D. Gershman, and B. Giles (2019), Impulsively Reflected Ions: A Plausible Mechanism for Ion Acoustic Wave Growth in Collisionless Shocks, *J. Geophys. Res. Space Physics*, 124, 3, 1855-1865, doi:10.1029/2018JA026436

Greenstadt, E. W. (1985), Oblique, Parallel, and Quasi-Parallel Morphology of Collisionless Shocks, in *Collisionless Shocks in the Heliosphere: Reviews of Current Research, Geophys. Monogr. Ser.*, vol. 35, edited by B. T. Tsurutani and R. G. Stone, pp. 169-184, AGU, Washington, D. C., doi:10.1029/GM035p0169.

Greenstadt, E. W., F. L. Scarf, C. T. Russell, V. Formisano, and M. Neugebauer (1975), Structure of the quasi-perpendicular laminar bow shock, *J. Geophys. Res.*, 80, 502-514, doi:10.1029/JA080i004p00502

Gurnett, D.A., F.M. Neubauer, and R. Schwenn (1979a), Plasma wave turbulence associated with an interplanetary shock, *J. Geophys. Res.*, 84, 541-552, doi:10.1029/JA084iA02p00541

Gurnett, D. A., E. Marsch, W. Pilipp, R. Schwenn, and H. Rosenbauer (1979b), Ion acoustic waves and related plasma observations in the solar wind, *J. Geophys. Res.*, 84, 2029-2038, doi:10.1029/JA084iA05p02029

Hanson, E. L. M., O. V. Agapitov, F. S. Mozer, V. Krasnoselskikh, S. D. Bale, L. Avanov, Y. Khotyaintsev, and B. Giles (2019), Cross-shock potential in rippled vs planar quasi-perpendicular shocks observed by MMS, *Geophys. Res. Lett.*, 46, 2381-2389. https://doi.org/10.1029/2018GL080240

Hellinger, P., P. Trávníček and J. D. Menietti (2004), Effective collision frequency due to ion-acoustic instability: theory and simulations, Geophys. Res. Lett., 31, L10806, doi:10.1029/2004GL020028

Hess, R. A., R. J. MacDowall, B. Goldstein, M. Neugebauer, and R. J. Forsyth (1998), Ion acoustic-like waves observed by Ulysses near interplanetary shock waves in the three-dimensional heliosphere, *J. Geophys. Res.*, 103, 6531-6541, doi:10.1029/97JA03395

Hobara, Y., M. Balikhin, V. Krasnoselskikh, M. Gedalin, and H. Yamagishi (2010), Statistical study of the quasi-perpendicular shock ramp widths, *J. Geophys. Res.*, 115, A11106, doi:10.1029/2010JA015659

Hoppe, M. M., C. T. Russell, L. A. Frank, T. E. Eastman, and E. W. Greenstadt (1981), Upstream hydromagnetic waves and their association with backstreaming ion populations: ISEE 1 and 2 observations, *J. Geophys. Res. Space Physics*, 86(A6), 4471-4492, doi:10.1029/ja086ia06p04471

Hoppe, M. M., C. T. Russell, T. E. Eastman, and L. A. Frank (1982), Characteristics of the ULF Waves Associated with Upstream Ion Beams, *J. Geophys. Res.*, 87, 643-650, doi:10.1029/JA087iA02p00643

Hoppe, M. M. and C. T. Russell (1983), Plasma Rest Frame Frequencies and Polarizations of the Low-Frequency Upstream Waves: ISEE 1 and 2 Observations, *J. Geophys. Res.*, 88, 2021-2028, doi:10.1029/JA088iA03p02021

Hull, A. J., D. E. Larson, M. Wilber, J. D. Scudder, F. S. Mozer, C. T. Russell, and S. D. Bale (2006), Large-amplitude electrostatic waves associated with magnetic ramp substructure at Earth's bow shock, *Geophys. Res. Lett.*, 35(15), doi:10.1029/2005GL025564

Hull, A. J., L. Muschietti, M. Oka, D. E. Larson, F. S. Mozer, C. C. Chaston, J. W. Bonnell, and G. B. Hospodarsky (2012), Multiscale whistler waves within Earth's perpendicular bow shock, *J. Geophys. Res. Space Physics*, 117 (A12), doi:10.1029/2012ja017870

Hutchinson, I. H. and D. M. Malaspina (2018), Prediction and Observation of Electron Instabilities and Phase Space Holes Concentrated in the Lunar Plasma Wake, *Geophys. Res. Lett.*, 45, 3838-3845, doi:10.1029/2017GL076880

Kajdič, P., X. Blanco-Cano, E. Aguilar-Rodriguez, C. T. Russell, L. K. Jian and J. G. Luhmann (2012), Waves Upstream and Downstream of Interplanetary Shocks Driven by Coronal Mass Ejections, *J. Geophys. Res.*, 117 (A6), doi:10.1029/2011JA017381

Kanekal, S. G. et al. (2016), Prompt acceleration of magnetospheric electrons to ultrarelativistic energies by the 17 March 2015 interplanetary shock*, J. Geophys. Res: Space Physics*, 121(8), 7622-7635, doi:10.1002/2016ja022596

Kellogg, P. J. (2003), Langmuir waves associated with collisionless shocks; a review, *Planetary and Space Science*, 51, 681-691, doi:10.1016/j.pss.2003.05.001

Kennel, C.F. (1987), Critical Mach numbers in classical magnetohydrodynamics, *J. Geophys. Res.*, 92, 13,427-13,437, doi:10.1029/JA092iA12p13427

Kennel, C. F. and R. Z. Sagdeev (1967a), Collisionless shock waves in high beta plasmas, 1, *J. Geophys. Res.*, 72, 3303- 3326, doi:10.1029/JZ072i013p03303

Kennel, C. F. and R. Z. Sagdeev (1967b), Collisionless shock waves in high beta plasmas, 2, *J. Geophys. Res.*, 72, 3327- 3341, doi:10.1029/JZ072i013p03327

Kennel, C.F., J.P. Edmiston, and T. Hada (1985), A quarter century of collisionless shock research, in *Collisionless Shocks in the Heliosphere: A Tutorial Review, Geophys. Monogr. Ser.*, vol. 34, edited by R.G. Stone and B.T. Tsurutani, pp. 1-36, AGU, Washington, D.C., doi:10.1029/GM034p0001

Krasnoselskikh, V., M. Balikhin, S. N. Walker, S. Schwartz, D. Sundkvist, V. Lobzin, M.
Gedalin, S. D. Bale, F. Mozer, J. Soucek, Y. Hobara and H. Comisel, The Dynamic
Quasiperpendicular Shock: Cluster Discoveries, Microphysics of Cosmic Plasmas, 10.1007/978-
1-4899-7413-6_18, (459-522), (2013).

Koval, A. and A. Szabo (2008), Modified ''Rankine-Hugoniot'' shock fitting technique: Simultaneous solution for shock normal and speed, *J. Geophys. Res.*, 113, A10110, doi:10.1029/2008JA013337

Koval, A., and A. Szabo (2010), Multispacecraft observations of interplanetary shock shapes on the scales of the Earth's magnetosphere*, J. Geophys. Res.: Space Physics,* 115(A12), doi:10.1029/2010ja015373

Lacombe, C., O. Alexandrova, L. Matteini, and O. Santolík (2014), Whistler mode waves and the electron heat flux in the solar wind: CLUSTER observations, *The Astrophys. J.*, 796:5, doi:10.1088/0004-637X/796/1/5

Lin, N., P. J. Kellogg, R. J. MacDowell, E. E. Scime, A. Balogh, R. J. Forsyth, D. J. McComas, and J. L. Phillips (1998), Very low frequency waves in the heliosphere: Ulysses observations. *J. Geophys. Res.*, 103, 12,023-12,035. doi:10.1029/98JA00764

Lumme, E., E. K. J. Kilpua, A. Isavnin, and K. Andreeova (2017), Database of Heliospheric Shock Waves: Method Documentation, http://www.ipshocks.fi/documentation

MacDonald, G. J. (1989), Spectral Analysis of Time Series Generated by Nonlinear Processes, *Rev. Geophys.*, 27, 449-469, doi:10.1029/RG027i004p00449

Matsumoto, H., et al. (1994), Electrostatic Solitary Waves (ESW) in the magnetotail: BEN wave forms observed by GEOTAIL, *Geophys. Res. Lett.*, 21, 2915-2918, doi:10.1029/94GL01284

Matsumoto, H. and H. Usui (1997), Intense bursts of electron cyclotron harmonic waves near the dayside magnetopause observed by GEOTAIL, *Geophys. Res. Lett.*, 24, 1, 49-52, doi:10.1029/96GL03650

Mazelle, C., Lembège, B., Morgenthaler, A., Meziane, K., Horbury, T. S., Génot, V., Lucek, E. A., and Dandouras, I. (2010). Self-reformation of the quasi-perpendicular shock: CLUSTER observations. *AIP Conference Proceedings*, 1216, 471 (2010). doi:10.1063/1.3395905

McFadden, J. P., et al. (2008), The THEMIS ESA Plasma Instrument and In-flight Calibration, *Space Sci. Rev.*, 141, 277-302, doi:10.1007/s11214-008-9440-2

Mean, J. D. (1972), Use of the three-dimensional covariance matrix in analyzing the polarization properties of plane waves, *J. Geophys. Res.*, 77, 28, 5551-5559, doi:10.1029/JA077i028p05551

Mellot, M. M. (1985), Subcritical Collisionless Shock Waves, in *Collisionless Shocks in the Heliosphere: Reviews of Current Research, Geophys. Monogr. Ser.*, vol. 35, edited by B. T. Tsurutani and R. G. Stone, pp. 131-140, AGU, Washington, D. C., doi:10.1029/GM035p0131.

Mellott, M. M. and E. W. Greenstadt (1984), The structure of oblique subcritical bow shocks: ISEE 1 and 2 observations, *J. Geophys. Res.*, 89, 2151-2161, doi:10.1029/JA089iA04p02151

Morlet, J. (1982), Wave propagation and sampling theory Part II: Sampling theory and complex waves, *J. Geophys. Res.*, 47, 222-236, doi:10.1190/1.1441329

Morlet, J., G. Arens, I. Forgeau, and D. Giard (1982), Wave propagation and sampling theory Part I: Complex signal and scattering in multilayered media, *J. Geophys. Res.*, 47, 203-221, doi:10.1190/1.1441328

Möstl, C. et al. (2012), Multi-point shock and flux rope analysis of multiple interplanetary coronal mass ejections around 2010 august 1 in the inner heliosphere, *Astro. J.*, 758(1), 10, doi:10.1088/0004-637X/758/1/10

Mozer, F. S., S. D. Bale, J. W. Bonnell, C. C. Chaston, I. Roth, and J. Wygant (2013), Megavolt Parallel Potentials Arising from Double-Layer Streams in the Earth's Outer Radiation Belt, *Phys. Rev. Lett.*, 111(23), 235002, doi:10.1103/PhysRevLett.111.235002

Neugebauer, M., and J. Giacalone (2005), Multispacecraft observations of interplanetary shocks: Nonplanarity and energetic particles, *J. Geophys. Res.*, 110, A12106, doi:10.1029/2005JA011380

Oliveira, D.M., Samsonov, A.A., Geoeffectiveness of interplanetary shocks controlled by impact angles: A review, Advances in Space Research (2017), doi:https://doi.org/10.1016/j.asr.2017.10.006

Papadopoulos, K. (1985), Microinstabilities and anomalous transport, in *Collisionless Shocks in the Heliosphere: A Tutorial Review, Geophys. Monogr. Ser.*, vol. 34, edited by R. G. Stone and B. T. Tsurutani, pp. 59– 90, AGU, Washington, D. C., doi:10.1029/GM034p0059

Petkaki, P. and M.P. Freeman (2008), Nonlinear dependence of anomalous ion-acoustic resistivity on electron drift velocity, *Astrophys. J.*, 686, 686–693, doi:10.1086/590654

Petkaki, P., M.P. Freeman, T. Kirk, C.E.J. Watt, and R.B. Horne (2006), Anomalous resistivity and the nonlinear evolution of the ion-acoustic instability, *J. Geophys. Res.*, 111, A01205, doi:10.1029/2004JA010793

Pope, S. A., M. Gedalin, and M. A. Balikhin (2019), The first direct observational confirmation of kinematic collisionless relaxation in very low mach number shocks near the Earth, *J. Geophys. Res.: Space Physics*, 124. 1711-1725, doi:10.1029/2018JA026223

Pulupa, M. and S. D. Bale (2008), Structure on interplanetary shock fronts: Type II radio burst source regions, *Astrophys. J.*, 676(2), 1330-1337, doi:10.1086/526405

Ramírez Vélez, J. C., X. Blanco-Cano, E. Aguilar-Rodriguez, C. T. Russell, P. Kajdič, L. K. Jian and J. G. Luhmann (2012), Whistler Waves Associated with Weak Interplanetary Shocks, *J. Geophys. Res.*, 117 (A11103), doi:10.1029/2012JA017573

Rodriguez, P. and D. A. Gurnett (1975), electrostatic and electromagnetic turbulence associated with the Earth's bow shock, *J. Geophys. Res.*, 80(1), 19-31, doi:10.1029/JA080i001p00019

Roux, A., O. Le Contel, C. Coillot, A. Bouabdellah, B. de la Porte, D. Alison, S. Ruocco, and M. C. Vassal (2008), The Search Coil Magnetometer for THEMIS, *Space Sci. Rev.*, 141, 265-275, doi:10.1007/s11214-008-9455-8

Russell, C. T. and M. M. Hoppe (1983), Upstream Waves and Particles, *Space Sci. Rev.*, 34, 155-172

Russell, C. T., L. K. Jian, X. Blanco-Cano and J. G. Luhmann (2009), STEREO Observations of Upstream and Downstream Waves at Low Mach Number Shocks, *Geophys. Res. Lett.*, 36 (3), doi:10.1029/2008GL036991

Sagdeev, R. Z. (1966), Cooperative phenomena and shock waves in collisionless plasmas, *Rev. Plasma Phys.*, 4, 23- 91

Samson, J. C. and J. V. Olson (1980), Some comments on the descriptions of the polarization states of waves, *Geophys. J. R. astr. Soc.*, 61, 115-129, doi:10.1111/j.1365-246X.1980.tb04308.x

Scholer, M. and J. W. Belcher (1971), The effect of Alfven waves on MHD fast shocks. *Solar Physics*, 16(2), 472-483. doi:10.1007/bf00162490

Schwartz, S. J. and D. Burgess (1991), Quasi-parallel shocks: A patchwork of three-dimensional structures, *Geophys. Res. Lett.*, 18(3), 373-376, doi:10.1029/91gl00138

Scudder, J.D., T.L. Aggson, A. Mangeney, C. Lacombe, and C.C. Harvey (1986a), The resolved layer of a collisionless, high beta, supercritical, quasi-perpendicular shock wave: I. Rankine-Hugoniot geometry, currents, and

stationarity, *J. Geophys. Res.*, 91, 11,019–11,052, doi:10.1029/JA091iA10p11019

Scudder, J.D., T.L. Aggson, A. Mangeney, C. Lacombe, and C.C. Harvey (1986b), The resolved layer of a collisionless, high beta, supercritical, quasi-perpendicular shock wave: II. Dissipative fluid electrodynamics, *J. Geophys. Res.*, 91, 11,053–11,073, doi:10.1029/JA091iA10p11053

Scudder, J.D., A. Mangeney, C. Lacombe, C.C. Harvey, and C.S. Wu (1986c), The resolved layer of a collisionless, high beta, supercritical, quasi-perpendicular shock wave: III. Vlasov electrodynamics, *J. Geophys. Res.*, 91, 11,075–11,097, doi:10.1029/JA091iA10p11075

Su, Y., Q. Lu, X. Gao, C. Huang, and S. Wang (2012), Ion dynamics at supercritical quasi-parallel shocks: Hybrid simulations, *Phys. Plasmas*, 19(9), 092,108, doi:10.1063/1.4752219

Szabo, A. (2005), Multi-spacecraft observations of interplanetary shocks, *AIP Conference Proceedings*, doi:10.1063/1.2032672

Terasawa, T., et al. (2005), Determination of shock parameters for the very fast interplanetary shock on 29 October 2003, *J. Geophys. Res.*, 110, A09S12, doi:10.1029/2004JA010941

Tidman, D. A. and N. A. Krall (1971), *Shock Waves in Collisionless Plasmas*, John Wiley, New York

Tidman, D. A. and T. G. Northrop (1968), Emission of plasma waves by the Earth's bow shock, *J. Geophys. Res.*, 73(5), 1543–1553. doi:10.1029/ja073i005p01543

Thomsen, M. F., J. T. Gosling, S. J. Bame, and M. M. Mellott (1985), Ion and electron heating at collisionless shocks near the critical Mach number, *J. Geophys. Res.*, 90, 137– 148, doi:10.1029/JA090iA01p00137

Tong, Y., I. Y. Vasko, A. V. Artemyev, S. D. Bale, and F. S. Mozer (2019), Statistical study of whistler waves in the solar wind at 1 au, *The Astrophys. J.*, 878:41, doi:10.3847/1538-4357/ab1f05

Usui, H., W. R. Paterson, H. Matsumoto, L. A. Frank, M. Nakamura, H. Matsui, T. Yamamoto, O. Nishimura, and J. Koizumi (1999), Geotail electron observations in association with intense burst of electron cyclotron harmonic waves in the dayside magnetosphere, *J. Geophys. Res.*, 104, 43, 4477-4484, doi:10.1029/1998JA900151

Wang, R., et al. (2020), Electrostatic Turbulence and Debye-scale Structures in Collisionless Shocks, *The Astrophys. J.*, 889, L9, doi:10.3847/2041-8213/ab6582

Watt, C. E. J., R. B. Horne, and M. P. Freeman (2002), Ion-acoustic resistivity in plasma with similar ion and electron temperatures, *Geophys. Res. Lett.*, 29, 1004, doi:10.1029/2001GL013451

Williams, J. D., L.-J. Chen, W. S. Kurth, D. A. Gurnett, M. K. Dougherty, and A. M. Rymer (2005), Electrostatic solitary structures associated with the November 10, 2003, interplanetary shock at 8.7 AU, *Geophys. Res. Lett.*, 32, L17103, doi:10.1029/2005GL023079

Wilson III, L. B., C. Cattell, P. J. Kellpgg, K. Goetz, K. Kersten, L. Hanson, R. MacGregor, and J. C. Kasper (2007), Waves in Interplanetary Shocks: A Wind/WAVES Study, *Phys. Rev. Lett.*, 99, 041101, doi: 10.1103/PhysRevLett.99.041101

Wilson III, L. B., et al. (2010), Large-amplitude electrostatic waves observed at a supercritical interplanetary shock, *J. Geophys. Res.*, 115, A12104, doi:10.1029/2010JA015332

Wilson III, L. B., et al. (2012), Observations of electromagnetic whistler precursors at supercritical interplanetary shocks, *Geophys. Res. Lett.*, 39, L08109, doi:10.1029/2012GL051581

Wilson III, L. B., et al. (2013). Electromagnetic waves and electron anisotropies downstream of supercritical interplanetary shocks. *J. Geophy. Res. Space Physics*,118, 5- 16. doi:10.1029/2012JA018167

Wilson III, L. B., et al. (2014a), Quantified energy dissipation rates in the terrestrial bow shock: 1. Analysis techniques and methodology, *J. Geophys. Res. Space Physics*, 119, 6455- 6474, doi:10.1002/2014JA019929

Wilson III, L. B., et al. (2014b), Quantified energy dissipation rates in the terrestrial bow shock: 2. Waves and dissipation, *J. Geophys. Res. Space Physics*, 119, 6475- 6495, doi:10.1002/2014JA019930

Wilson, L. B. (2016), Low Frequency Waves at and Upstream of Collisionless Shocks*, Geophys. Monograph Series*, 269-291, doi:10.1002/9781119055006.ch16

Wilson III, L. B., A. Koval, A. Szabo, M. L. Stevens, J. C. Kasper, C. A. Cattell, and V. V. Krasnoselskikh (2017), Revisiting the structure of low-Mach number, low-beta, quasi-perpendicular shocks, *J. Geophys. Res.*, 122, 9115-9133, doi:10.1002/2017JA024352

Wygant, J. R., et al. (2013), The electric field and waves instruments on the radiation belt storm probes mission, *Space Sci. Rev.*, 179, 183- 220, doi:10.1007/s11214-013-0013-7

Yoon, P.H. and A.T.Y. Lui (2006), Quasi-linear theory of anomalous resistivity, *J. Geophys. Res.*,111, A02203, doi:10.1029/2005JA011482

Yoon, P.H. and A.T.Y. Lui (2007), Anomalous resistivity by fluctuation in the lower-hybrid frequency range, *J. Geophys. Res.*, 112, A06207, doi:10.1029/2006JA012209